\documentclass[useAMS,usenatbib,usegraphicx]{mn2e}
\usepackage{xspace,amsmath}
\usepackage{color,epsfig,amssymb,lscape}
\usepackage{util}
\usepackage{array,colortbl}
\usepackage[colorlinks=true,linkcolor=magenta,citecolor=blue,breaklinks=true]{hyperref}
\usepackage{ifpdf}

\ifpdf
\usepackage{epstopdf}
\else
\usepackage[dvips]{graphicx}
\usepackage{lscape}
\fi

\title[Metallicity and Star Formation History of NGC 6789]
{Auto-consistent metallicity and star formation history of the nearest 
blue compact dwarf galaxy NGC 6789}
\author[R. Garc\'ia-Benito \& E. P\'erez-Montero]
{R. Garc\'ia-Benito$^{1,2}$\thanks{E-mail:
rgb@iaa.es} and E. P\'erez-Montero$^{2}$ \\
$^{1}$Kavli Institute of Astronomy and Astrophysics, Peking University, 100871, Beijing, China\\
$^{2}$Instituto de Astrof\'isica de Andaluc\'ia, CSIC, Apartado de correos 3004, 18080 Granada, Spain\\
}

\begin{document}

\date{Accepted 2012 March 8}

\pagerange{\pageref{firstpage}--\pageref{lastpage}} \pubyear{2012}

\maketitle

\label{firstpage}

\begin{abstract}
We present a detailed auto-consistent study of the nearest blue compact dwarf galaxy NGC 6789
by means of optical and UV archive photometry data and optical long-slit ISIS-WHT spectroscopy 
observations of the five brightest star-forming knots. The analysis of the spectra in all knots 
allowed the derivation of ionic chemical abundances of oxygen, nitrogen, sulphur, argon and neon
using measures of both the high- and low-excitation electron temperatures, leading to the  
conclusion that NGC 6789 is chemically homogeneous with low values of the abundance of oxygen 
in the range 12+log(O/H) = 7.80-7.93, but presenting at the same time higher values of the 
nitrogen-to-oxygen ratio than expected for its metal regime.

We used archival \textit{Hubble Space Telescope}/Wide-Field Planetary Camera 2 (HST/WFPC2) 
F555W and F814W observations of NGC 6789 to perform a photometric study of the colour-magnitude 
diagram (CMD) of the resolved stellar populations and derive its star formation history (SFH),
which is compatible with the presence of different young and old stellar populations 
whose metallicities do not necessarily increase with age.
We fit the observed optical spectrum in all the five knots using the STARLIGHT code 
and a combination of single stellar populations following the SFH obtained from the CMD.
We compare the resulting stellar masses and the relative fractions of the
ionising populations with a non-constrained SFH case.
The properties of the younger populations were obtained using CLOUDY
photoionisation models, giving similar ages in all the knots in the range 3-6 Myr and 
the estimation of the dust absorption factor, which correlates with the observed GALEX 
FUV-NUV colour indices. The total photometric extinction and dust-absorption corrected 
H$\alpha$ fluxes were finally used to derive the star formation rates.

\end{abstract}

\begin{keywords}
ISM: abundances -- H\,{\sc ii} regions -- galaxies: dwarf -- galaxies: individual: NGC 6789 -- 
galaxies: starburst -- galaxies: star formation -- Hertzsprung-Russell and colour-magnitude diagrams
\end{keywords}

\section{Introduction}

Blue compact dwarf galaxies (BCDs) are known by their intense processes of star-formation. 
These make that their optical spectra were dominated by the blue light from massive young 
stars and the bright emission lines from the ionised gas, so they are also called
H {\sc ii} galaxies.

Their high specific star-formation rates and their low gas-phase metallicities were taken as 
evidences that these galaxies were undergoing  their first bursts of star formation 
\citep{sargent70}. In fact, one of the most important aspects about BCDs is that they 
constitute an important link to the high redshift universe and the early epoch of galaxy 
formation \citep{bergvall02}. During the last decade, however, deep imaging of BCDs showed 
that in addition to bright young stars from the present starburst most of them have an underlying 
older stellar population \citep{papaderos96a,aloisi99,ostlin00}. 

One of the challenges of the BCDs is to reconcile their low--observed metallicity with the 
relatively high star formation rate (SFR) in these galaxies. \cite{matteucci83} proposed 
three possible mechanisms to explain this fact: 1) variations in the initial mass function 
(IMF); 2) accretion of metal-poor gas; and 3) galactic winds powered by supernovae explosions. 
According to numerical models, it seems that the last mechanism is the most simple to reproduce 
the observed properties (see \citealt{tolstoy09} and references therein for a detailed 
explanation).

The star formation history (SFH) and the metal content of BCDs must be figured out in order 
to shed some light on these questions. In the case of SFH, although some advances have been 
done by means of the fitting of the optical spectrum with stellar populations synthesis  
\citep{perez-montero10}, one of the most reliable method consists of the analysis of 
colour-magnitude diagrams (CMDs). However, this is limited to the closest objects whose stellar 
population can be resolved. Another important contribution comes from the study of the ionic 
chemical abundances of elements with a different nucleosynthetic origin, such as  
oxygen and nitrogen \citep{molla06}. These chemical abundances can be derived very accurately 
in the metal-poor gas phase of BCDs, where the cooling rate is not efficient, increasing the 
electron temperature of the ionised gas and enhancing the emissivity of the collisional lines 
necessary to derive the ionic abundances following the method based on the determination of 
the electron temperature \citep{pmontero03,hagele06}.

NGC\,6789 is, according to \cite{drozdovsky01}, the closest BCD to our Galaxy, with a distance 
modulus of (m - M) = 27.80, or a distance of 3.6 Mpc. This value yields a linear scale of 
17.5 pc arcsec$^{-1}$. \cite{drozdovsky00} imaged NGC 6789 from the ground. They found that 
it belongs to the iE subtype (following \citealt{loose86} classification), exhibiting the 
morphology most characteristic of the vast majority of BCDs. NGC 6789 presents several H$\alpha$ 
emission knots which show evidence of actual star formation activity. It also shows a high 
surface brightness in its central region and a small radial velocity 
\citep[V$_0$ = -141 km s$^{-1}$,][]{karachentsev98}. NGC\, 6789's closeness, spatial isolation 
and morphology offer the prospect of studying the structure, SFH, and metallicity of different 
star forming knots within the same BCD. 

To investigate these issues, we obtained simultaneous blue and red long-slit observations 
with the ISIS double-arm spectrograph at the 4.2m William Herschel Telescope (WHT), 
of the brightest five knots of NGC 6789. We also used archival data from HST/WFPC2, 
GALEX and H$\alpha$ narrow filter, and B and R broad filters. These images were used to
derive the SFH and to give an observational input to derive the properties of the
ionising populations with the aid of photoionisation models. These were
also used to provide more accurate chemical abundances in each one of the five 
studied knots of this galaxy and to better understand their differential 
chemical evolution.

In the following section the long-slit WHT observations and reduction are described 
and the results from the analysis are presented. In section \ref{phot} the optical and UV 
photometry are described together with the resolved stellar photometry. Finally, we discuss 
all these results in Section \ref{discussion} and conclusions are presented in 
Section \ref{conclusions}.

\section{Long-slit spectroscopy}
\label{spec}

\subsection{Observations and reduction}

The long-slit spectrophotometric observations of NGC~6789 were obtained using the 
ISIS double-beam spectrograph mounted on the 4.2 m William Herschel Telescope (WHT) of the
Isaac Newton group (ING) at the Roque de los Muchachos Observatory on the Spanish 
island of La Palma. They were acquired on 2005 July 8 during one single night observing run
and under photometric conditions, with an average seeing of 0.7 arcsec. The EEV12 and 
Marconi2 detectors were attached to the blue and red arms of the spectrograph, respectively. 
The R600B grating was used in the blue covering the wavelength range 3670 - 5070 \AA\ 
(centred at $\lambda_c$ = 4370 \AA), giving a spectral dispersion of 0.45 \AA\ pixel$^{-1}$. 
On the red arm, the R316R grating was mounted in two different central wavelengths providing
a spectral range from 5500 to 7800 \AA\ ($\lambda_c$ = 6650 \AA) and
from 7600 to 9900 \AA\ ($\lambda_c$ = 8750 \AA) with a spectral dispersion
of 0.86 \AA\ pixel$^{-1}$. To reduce the readout noise of our images,
the observations were taken with the ``SLOW'' CCD speed. The
pixel size for this set-up is 0.2 arcsec for both spectral
ranges. The slit width was 1 arcsec, which, combined with
the spectral dispersions, yields spectral resolutions of about 1.0
and 3.5 \AA\ FWHM in the blue and red arms, respectively. 
The instrumental configuration and other details on the exposures 
are given in the journal of observations in Table \ref{journal}. 

\begin{table*}
\centering
\caption[]{WHT instrumental configuration}
\label{journal}
\begin{tabular} {l c c c c c c}
\hline
Slit position &  Spectral range  &       Disp.             & FWHM   & Spatial res.         & Exposure Time    \\
              & (\AA)          & (\AA\,px$^{-1}$)        & (\AA)  & (\arcsec\,px$^{-1}$) & s                \\
\hline
S1 & 3670-5070       &       0.45              &  1.0   &   0.2                & 4 $\times$ 900                  \\
S1 & 5500-7800       &       0.86              &  3.5   &   0.2                & 2 $\times$ 900, 1 $\times$ 300  \\
S1 & 7600-9900       &       0.86              &  3.5   &   0.2                & 2 $\times$ 900                \\
S2 & 3670-5070       &       0.45              &  1.0   &   0.2                & 5 $\times$ 900  \\
S2 & 5500-7800       &       0.86              &  3.5   &   0.2                & 2 $\times$ 900, 1 $\times$ 300  \\
S2 & 7600-9900       &       0.86              &  3.5   &   0.2                & 2 $\times$ 900  \\
\hline
\end{tabular}
\end{table*}

\begin{figure*}
\includegraphics[width=\linewidth,clip=]{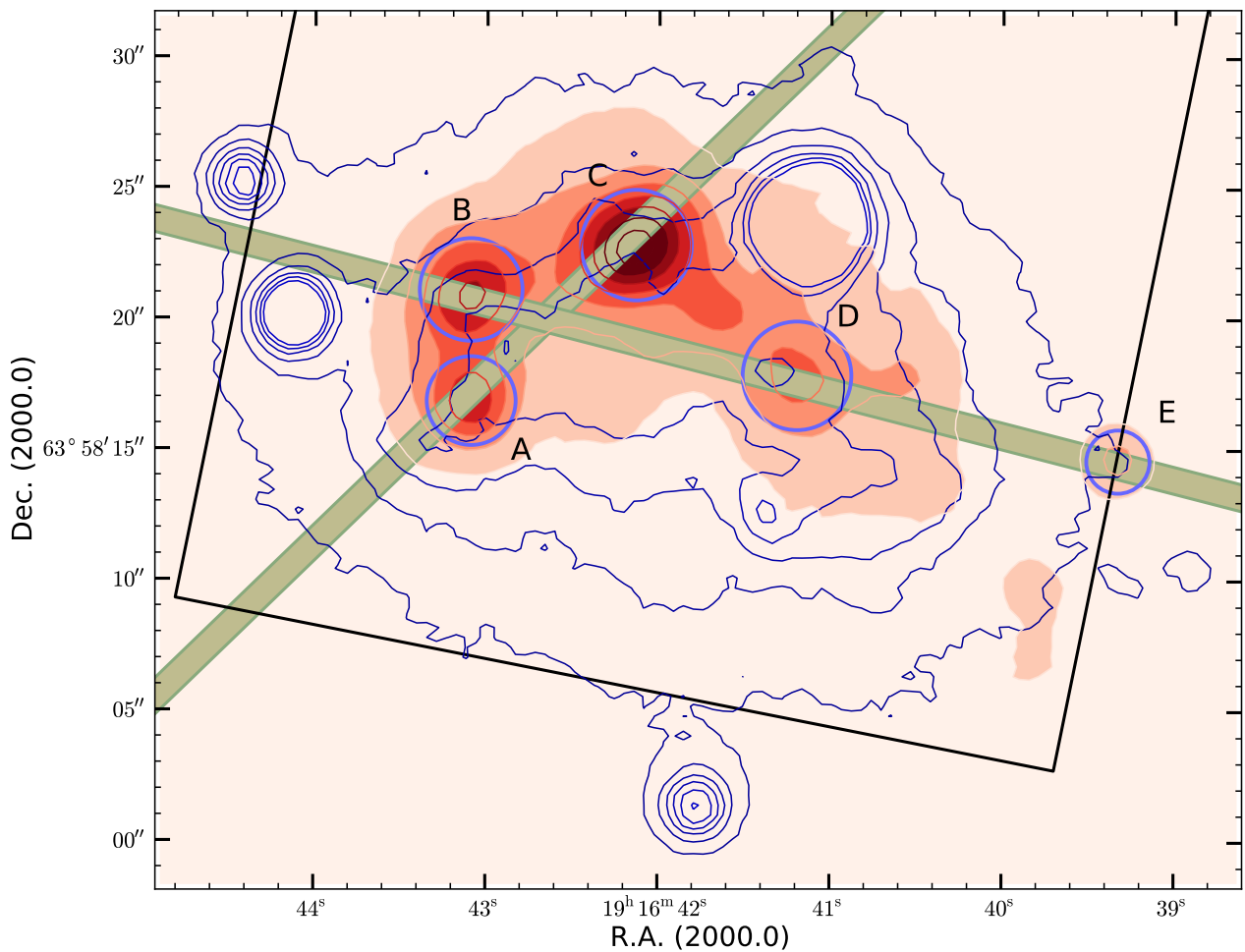}
\caption[]{H$\alpha$ image of NGC 6789 from \cite{gildepaz03} with the identification of the 
observed knots (labelled from A to E). We show the regions for which the H$\alpha$ flux was 
measured and the position of the two slits for the observations described in the text. The 
R-band contours show the position of the host galaxy. Finally, we also plot part of the rectangle 
encompassing the field-of-view of the PC chip of the WFPC2 data used to obtain the resolved stellar 
photometry described in Section \ref{rphot}. North is up and east is towards the left-hand side.}
\label{regslit}
\end{figure*}

Several bias and sky flat field frames were taken at the beginning and at the end
of the night in both arms. In addition, two lamp flat fields and one calibration
lamp exposure were taken for each telescope position. The calibration lamp
used was +CuAr. 

The images were processed and analysed with
IRAF\footnote{IRAF: the Image Reduction and Analysis Facility is distributed by
  the National Optical Astronomy Observatories, which is operated by the
  Association of Universities for Research in Astronomy, In. (AURA) under
  cooperative agreement with the National Science Foundation (NSF).} routines in
the usual manner. The procedure included cosmic rays removal, bias subtraction, 
division by a normalised flat field and wavelength calibration. Typical wavelength 
fits to second to third order polynomials were performed using around 40 
lines in the blue and 20-25 lines in the red. These fits were done at 100 different 
locations along the slit in both arms (beam size of 10 pixels) obtaining rms residuals 
between $\sim$0.1 and $\sim$0.2\,pix. In the last step, the spectra were corrected for 
atmospheric extinction and flux calibrated. For both arms, BD+254655 standard star 
observations were used, allowing a good spectrophotometric calibration with an estimated 
accuracy of about 5\%.

Unfortunately, the sky subtraction in the spectral range from 7600 to 9900 \AA\ left 
strong residuals from night-sky emission lines and telluric absorption, leaving this 
part of the spectra unusable and therefore, we were not able to measure with 
enough accuracy the [S{\sc iii}]\,$\lambda\lambda$\,9069,9532\,\AA\ lines in any of 
the five knots.

Figure \ref{regslit} shows the H$\alpha$ image and continuum contours in the R-band
from \cite{gildepaz03} in order to illustrate the position of the slits in relation to the 
position of the bursts of star formation and the host galaxy. We also show elliptical 
regions in H$\alpha$ taken to measure the total H$\alpha$ and GALEX flux of the knots
(see Section \ref{Halfa} below). The five knots are labelled with the first alphabet letters. 
Knots A and C were taken at PA = 138$^o$, which in average is close to the parallactic angle. 
In the case of knots B, D, and E the slit was at PA = 76$^o$, which is in average 
$\approx$ 40$^o$ off the parallactic angle. 
Therefore, the observations in this second position can be affected by a
certain differential atmospheric refraction (DAR). Nevertheless, taking into
account the curves given by \cite{filippenko82} for the mean air mass during
the observations  ($\approx$ 1.30), we calculated that the angular deviation
between [O{\sc ii}] and [S{\sc ii}] is not larger than the slit width at this P.A.
Besides, we checked that the emission line-ratios in the spatial position
where both slit positions intersect do not vary in more than the quoted
errors. Nevertheless, the line measurements errors of the second 
slit position (B, D and E knots) might be larger due to this effect.
For the sake of comparison, we also plot in Figure \ref{regslit} part of the rectangle 
encompassing the field-of-view (FOV) of the PC chip of the WFPC2 data used to obtain the 
resolved stellar photometry (see Section \ref{rphot})


\subsection{Line intensities and reddening correction}
\label{line}

The optical calibrated spectra of the five observed knots with some relevant identified emission 
lines are shown in Figure \ref{spectra}. The spectrum of each knot is split into two panels.

\begin{figure*}
\includegraphics[width=\textwidth,clip=]{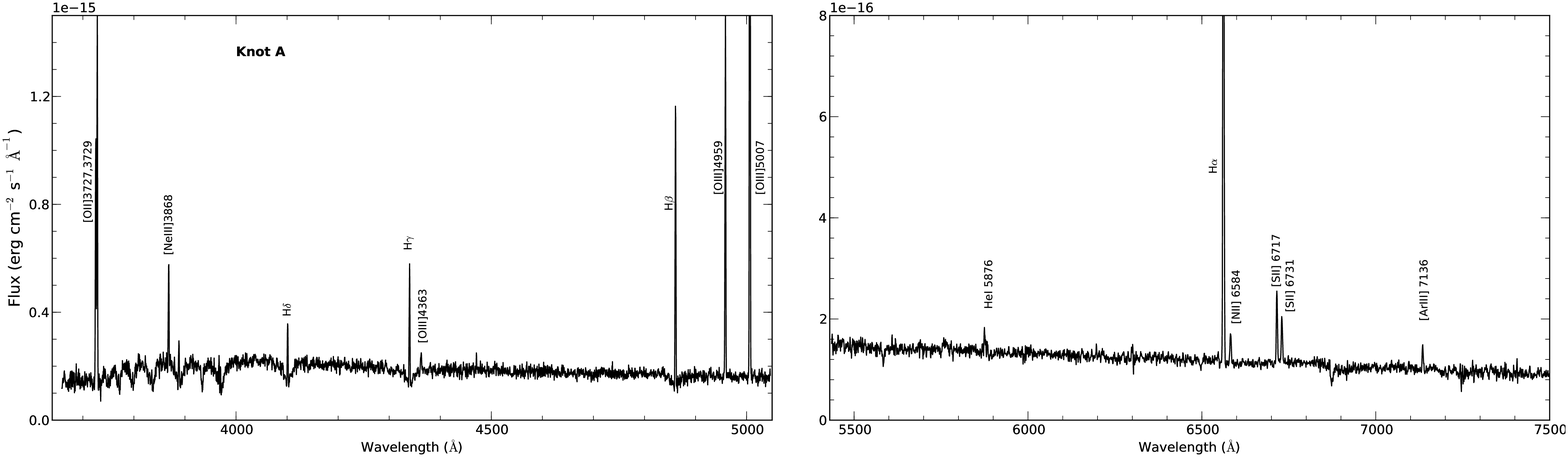}
\includegraphics[width=\textwidth,clip=]{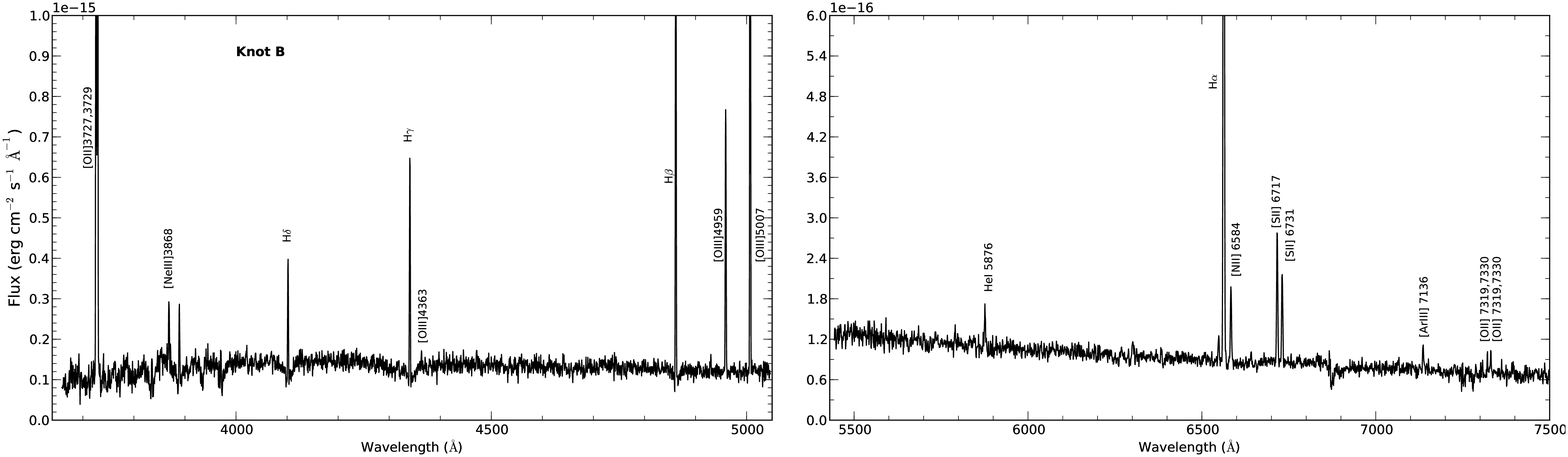}
\includegraphics[width=\textwidth,clip=]{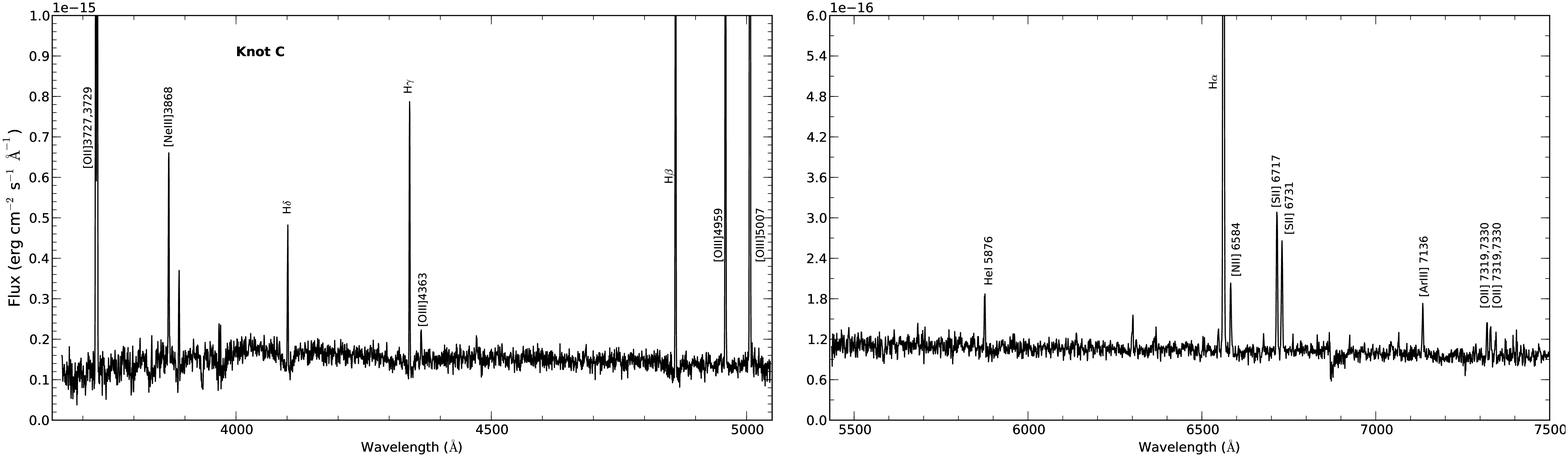}
\includegraphics[width=\textwidth,clip=]{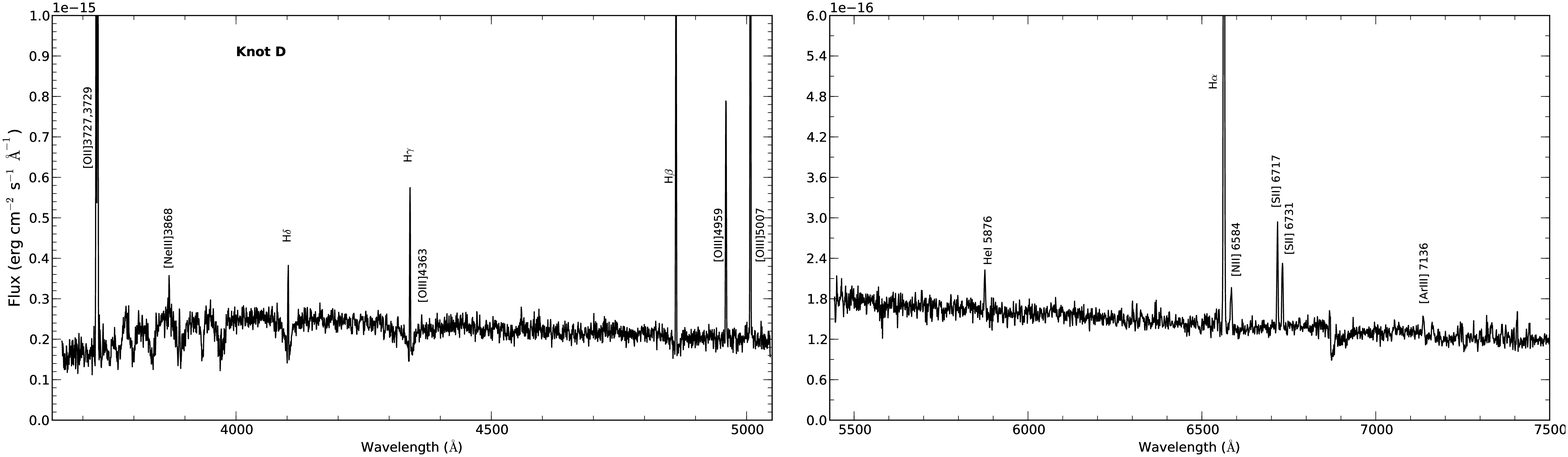}
\caption{Blue and red spectra for the knots A, B, C, D, and E of NGC~6789.}
\label{spectra}
\end{figure*}

\addtocounter{figure}{-1}
\begin{figure*}
\centering
\includegraphics[width=\textwidth,clip=]{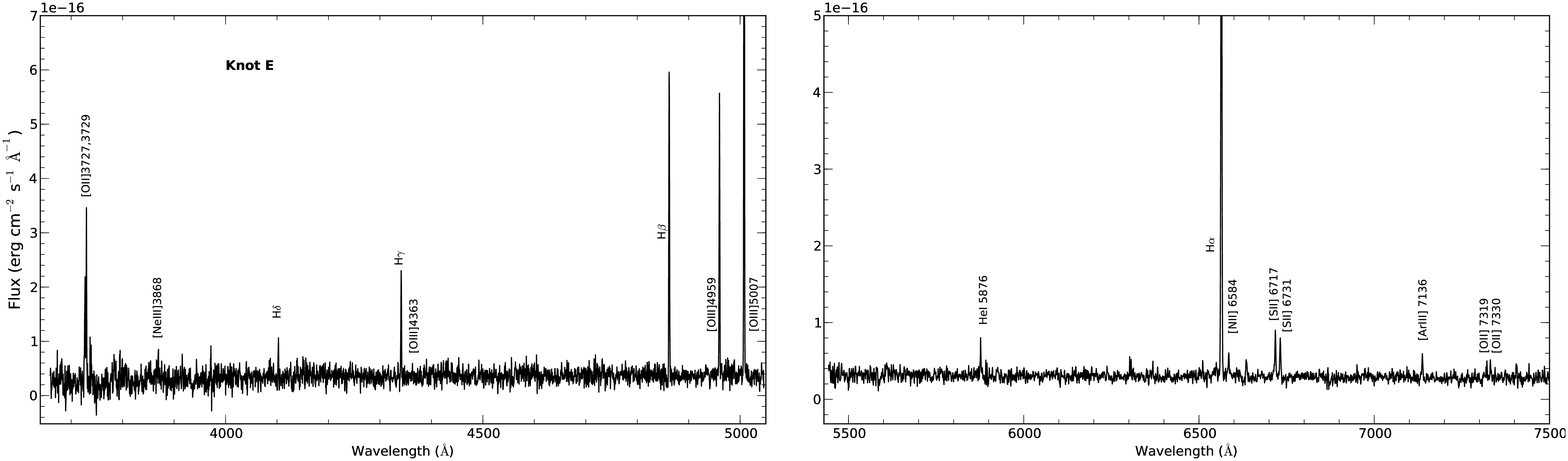}
\caption{-- {\it continued}}
\end{figure*}

Underlying stellar population in star forming galaxies have several effects in the measure of the 
emission lines produced by the ionised gas. Balmer and Paschen emission lines are depressed by the 
presence of absorption wings of stellar origin and does not allow the measure of their fluxes with 
acceptable accuracy \citep{diaz88}. All the properties derived from ratios that involve these 
lines, like reddening or ionic abundances, will be affected.

We subtracted from the observed spectra the spectral energy distribution of the underlying stellar 
population found by the spectral synthesis code STARLIGHT\footnote{The STARLIGHT project is 
supported by the Brazilian agencies CNPq, CAPES and FAPESP and by the France-Brazil CAPES/Cofecub 
program.} \citep{cidfernandes04,cidfernandes05,mateus06}. STARLIGHT fits an observed continuum 
spectral energy distribution using a combination of the synthesis spectra of different single 
stellar populations (SSPs; also known as instantaneous burst) using a $\chi^2$ minimisation 
procedure. We chose for our analysis the SSP spectra from \cite{bruzual03}, based on the STELIB 
library of \cite{leborgne03}, Padova 1994 evolutionary tracks, and a \cite{chabrier03} IMF between 
0.1 and 100 M$_{\odot}$. The metallicity of the libraries was strictly constrained following the 
results derived from Section \ref{ssfh}. Thus, we fixed the metallicity of the stellar populations to 
Z = 0.004 ($\approx$ 3/10 Z$_{\odot}$\footnote{Z$_\odot$ = 0.0122}) for the interval log(age) = 
[8.1,8.5] and  Z = 0.001 ($\approx$ 1/12 Z$_{\odot}$) for log(age) = \{6.0, 6.4, 6.6, 6.8, 7.0, 
7.1, 7.2, 7.9, 8.0, 8.1, 8.2, 8.3, 9.1, 9.2, 9.3, 9.7, 9.8, 9.9, 10.0, 10.1\} shown in the SFH solution 
(see Figure \ref{sfh}). As explained in section \ref{ssfh}, although both metallicities appear in 
the youngest burst ($<$ 12 Myr) SFH solution, we assigned the lower metallicity to this event 
according to the abundances derived in section \ref{abundances}.  
The STARLIGHT code solves simultaneously the ages and relative contributions of the different 
SSPs and the average reddening. The reddening law from \cite{cardelli89} with $R_{V}$ = 3.1 was 
used. Prior to the fitting procedure, the spectra were shifted to the rest frame and re-sampled 
to a wavelength interval of 1 \AA\ in the entire wavelength range by interpolation, as required by 
the program. Bad pixels and emission lines were excluded from the final fits.

It should be noted that while emission lines can be masked out, this is not possible for the 
nebular continuum emission. We tested the contribution of this emission using
Starburst99 libraries \citep[SB99;][]{leitherer99} in the same STARLIGHT models and we checked that 
the nebular continuum does not affect significantly neither the subtraction of the underlying continuum 
nor the determination of the stellar mass, consistently with the weak gas emission in this galaxy.

The H{\sc i} series (emitted as a consequence of recombination) were used to determine the extinction, 
comparing the observed line ratios with the expected theoretical values. Case B (optically thick in 
all the Lyman lines) is the best simple approximation to describe the physical conditions in the 
ionisation of the gas. This method takes advantage of the fact that the ratio between the emissivities 
of two hydrogen recombination lines, which depends on electron temperature and density, is almost 
constant. As an example, the ratio between the emissivity of H$\alpha$ and H$\beta$ is 2.86 for 
the case B with $n_{e} = 100$ $cm^{-3}$ and $T_{e} = 10000$ K, and this value varies less than 
10\% in the range of interest of temperatures and densities for an H{\sc ii} region.
We used an iterative method to estimate them, taking as starting values those derived from the 
measured [S{\sc ii}] $\lambda\lambda$ 6717,6731 \AA\ and 
[O{\sc iii}] $\lambda\lambda$ 4363,4959,5007 \AA. A least square fit of the measured decrements to 
the theoretical ones, computed based on the data by \cite{storey95}, was performed that provides 
the reddening coefficient, c(H$\beta$), and adopting the extinction law given by \cite{cardelli89} 
with R$_{V}$ = 3.1. Due to the large error introduced by the presence of the underlying population, 
only the strongest Balmer emission lines (H$\alpha$, H$\beta$, H$\gamma$, and H$\delta$) were used.

Line fluxes for the most relevant emission lines were measured using the \texttt{splot} task 
in {\sc IRAF}. Balmer lines were obtained from the STARLIGHT residual spectra, while the rest 
of the lines were measured on the original spectra. We checked in all spectra that the 
Starlight models properly reproduce the observed Balmer absorption profiles\footnote{From several 
fitting experiments using other libraries it is found that the resolution of the SSP spectra is a 
critical factor for Starlight in order to fit the absorption lines.}. The statistical errors 
associated with the observed emission line fluxes were calculated using the expression:

\begin{equation}
\sigma_{l} = \sigma_{c}\sqrt{N\left(1 + \frac{EW}{N\Delta}\right)}
\end{equation}

\noindent where $\sigma_{l}$ is the error in the observed line flux, $\sigma_{c}$ represents the 
standard deviation in a box near the measured emission line and stands for the error in the 
continuum placement, N is the number of pixels used in the measure of the line flux, EW is the line 
equivalent width, and $\Delta$ is the wavelength dispersion in \AA\ per pixel 
\citep{gdelgado94}. This expression takes into account the error in the continuum and the photon 
count statistics of the emission line. 

Table \ref{lines} gives the equivalent widths and the emission-line fluxes relative to 
1000$\cdot$F(H$\beta$), before and after reddening correction, in the optical spectra of the 
five observed knots, together with the reddening constants and their errors, their corresponding 
A(V) value\footnote{c(H$\beta$) = 0.4656*A(V), using \cite{cardelli89} and R$_{V}$ = 3.1.} and the 
extinction-corrected H$\beta$ flux. We also provide the adopted reddening curve, $f(\lambda)$ normalised 
to H$\beta$. The errors in the emission-line ratios were obtained by propagating in quadrature the 
observational errors in the emission-line fluxes and the reddening constant uncertainties.

\begin{table*}
\caption{Observed and reddening corrected relative line intensities [F(H$\beta)$=I(H$\beta)$=1000]
with their corresponding errors for the five knots. The adopted reddening curve, $f(\lambda)$ 
(normalised to H$\beta$), the equivalent width of the emission lines, the extinction-corrected 
H$\beta$ intensity, the reddening constant c(H$\beta$) and the corresponding A(V) are also given.}
\label{lines}
\begin{tabular}{lc@{\hspace{3em}}ccc@{\hspace{3em}}ccc}
\hline
\multicolumn{1}{l}{} & \multicolumn{1}{l}{} & \multicolumn{3}{c}{\hspace{-3em}\large NGC 6789-A} & \multicolumn{3}{c}{\large NGC 6789-B} \\
\multicolumn{1}{c}{$\lambda$ ({\AA})} & f($\lambda$) & F($\lambda$) & EW(\AA) & I($\lambda$) & F($\lambda$) & EW(\AA) & I($\lambda$) \\
\hline
 3726 [O{\sc ii}]             &     0.322  &       896 $\pm$    19 &    -15.0 &      979 $\pm$    63   &      1186 $\pm$    17 &    -34.9 &     1277 $\pm$    80  \\
 3729 [O{\sc ii}]             &     0.322  &      1241 $\pm$    16 &    -20.7 &     1357 $\pm$    84   &      1744 $\pm$    16 &    -51.9 &     1878 $\pm$   116  \\
 3835 H9                      &     0.299  &        37 $\pm$    10 &     -0.6 &       40 $\pm$    11   &        80 $\pm$     4 &     -3.5 &       85 $\pm$     7  \\
 3868 [Ne{\sc iii}]           &     0.291  &       302 $\pm$    17 &     -2.8 &      327 $\pm$    26   &       153 $\pm$    14 &     -3.7 &      164 $\pm$    17  \\
 3889 He{\sc i}+H8            &     0.286  &       153 $\pm$     6 &     -2.6 &      166 $\pm$    12   &       146 $\pm$    10 &     -4.5 &      156 $\pm$    14  \\
 3968 [Ne{\sc iii}]+H7        &     0.266  &        81 $\pm$     8 &     -1.3 &       88 $\pm$    10   &        $\cdots$       & $\cdots$ &         $\cdots$      \\
 4102 H$\delta$               &     0.229  &       247 $\pm$     5 &     -3.9 &      263 $\pm$    16   &       248 $\pm$    16 &     -6.7 &      261 $\pm$    22  \\
 4340 H$\gamma$               &     0.157  &       454 $\pm$    11 &     -7.2 &      474 $\pm$    28   &       448 $\pm$     7 &    -12.4 &      465 $\pm$    26  \\
 4363 [O{\sc iii}]            &     0.149  &        63 $\pm$     3 &     -0.7 &       66 $\pm$     5   &        23 $\pm$     4 &     -0.4 &       23 $\pm$     4  \\
 4471 He{\sc i}               &     0.115  &        36 $\pm$     4 &     -0.4 &       37 $\pm$     5   &        37 $\pm$     6 &     -0.7 &       39 $\pm$     7  \\
 4861 H$\beta$                &     0.000  &      1000 $\pm$    10 &    -17.3 &     1000 $\pm$    47   &      1000 $\pm$    12 &    -29.1 &     1000 $\pm$    48  \\
 4959 [O{\sc iii}]            &    -0.026  &      1240 $\pm$    12 &    -16.0 &     1231 $\pm$    57   &       512 $\pm$    12 &    -11.4 &      509 $\pm$    26  \\
 5007 [O{\sc iii}]            &    -0.038  &      3801 $\pm$    13 &    -50.1 &     3762 $\pm$   167   &      1492 $\pm$     9 &    -33.0 &     1479 $\pm$    66  \\
 5876 He{\sc i}               &    -0.203  &       102 $\pm$    14 &     -1.6 &       97 $\pm$    14   &        96 $\pm$    16 &     -2.3 &       92 $\pm$    16  \\
 6312 [S{\sc iii}]            &    -0.264  &        11 $\pm$     2 &     -0.2 &       10 $\pm$     2   &        16 $\pm$     5 &     -0.5 &       16 $\pm$     5  \\
 6548 [N{\sc ii}]             &    -0.296  &        28 $\pm$     5 &     -0.5 &       26 $\pm$     5   &        61 $\pm$     9 &     -1.8 &       57 $\pm$     8  \\
 6563 H$\alpha$               &    -0.298  &      3038 $\pm$    29 &    -57.0 &     2797 $\pm$    38   &      3003 $\pm$    15 &    -89.3 &     2807 $\pm$    20  \\
 6584 [N{\sc ii}]             &    -0.300  &       136 $\pm$    10 &     -2.5 &      126 $\pm$     9   &       186 $\pm$    11 &     -5.6 &      174 $\pm$    10  \\
 6717 [S{\sc ii}]             &    -0.318  &       277 $\pm$    10 &     -5.0 &      254 $\pm$    10   &       314 $\pm$     9 &     -9.5 &      292 $\pm$     8  \\
 6731 [S{\sc ii}]             &    -0.320  &       188 $\pm$    10 &     -3.4 &      172 $\pm$     9   &       217 $\pm$    10 &     -6.6 &      202 $\pm$     9  \\
 7136 [Ar{\sc iii}]           &    -0.374  &       112 $\pm$     7 &     -2.4 &      101 $\pm$     7   &        76 $\pm$     7 &     -2.7 &       70 $\pm$     7  \\
 7319 [O{\sc ii}]             &    -0.398  &        $\cdots$       & $\cdots$ &         $\cdots$       &        46 $\pm$     6 &     -1.6 &       42 $\pm$     6  \\
 7330 [O{\sc ii}]             &    -0.400  &        $\cdots$       & $\cdots$ &         $\cdots$       &        35 $\pm$     4 &     -1.2 &       32 $\pm$     4  \\
\hline
I(H$\beta$)(erg\,seg$^{-1}$\,cm$^{-2}$) & & \multicolumn{3}{c}{  2.70e-15} &  \multicolumn{3}{c}{  3.19e-15} \\
c(H$\beta$) & & \multicolumn{3}{c}{0.12 $\pm$ 0.02} &  \multicolumn{3}{c}{0.10 $\pm$ 0.02} \\
A(V) & & \multicolumn{3}{c}{0.26 $\pm$ 0.04} &  \multicolumn{3}{c}{0.21 $\pm$ 0.04} \\
\hline
\multicolumn{1}{l}{} & \multicolumn{1}{l}{} & \multicolumn{3}{c}{\hspace{-3em}\large NGC 6789-C} & \multicolumn{3}{c}{\large NGC 6789-D} \\
\multicolumn{1}{c}{$\lambda$ ({\AA})} & f($\lambda$) & F($\lambda$) & EW(\AA) & I($\lambda$) & F($\lambda$) & EW(\AA) & I($\lambda$) \\
\hline 
 3726 [O{\sc ii}]             &     0.322  &       940 $\pm$    13 &    -29.1 &     1020 $\pm$    64   &      1159 $\pm$    26 &    -15.4 &     1267 $\pm$    82  \\
 3729 [O{\sc ii}]             &     0.322  &      1377 $\pm$    12 &    -42.4 &     1494 $\pm$    92   &      1712 $\pm$    21 &    -22.5 &     1871 $\pm$   116  \\
 3835 H9                      &     0.299  &        47 $\pm$     9 &     -1.4 &       50 $\pm$    10   &        $\cdots$       & $\cdots$ &         $\cdots$      \\
 3868 [Ne{\sc iii}]           &     0.291  &       300 $\pm$    12 &     -5.2 &      323 $\pm$    23   &       105 $\pm$     9 &     -0.9 &      114 $\pm$    12  \\
 3889 He{\sc i}+H8            &     0.286  &       169 $\pm$    12 &     -5.5 &      182 $\pm$    17   &       167 $\pm$    20 &     -2.4 &      180 $\pm$    24  \\
 3968 [Ne{\sc iii}]+H7        &     0.266  &        74 $\pm$     9 &     -2.3 &       79 $\pm$    11   &        $\cdots$       & $\cdots$ &         $\cdots$      \\
 4102 H$\delta$               &     0.229  &       246 $\pm$    16 &     -5.9 &      261 $\pm$    22   &       244 $\pm$     7 &     -3.2 &      260 $\pm$    16  \\
 4340 H$\gamma$               &     0.157  &       446 $\pm$     7 &    -12.4 &      464 $\pm$    26   &       444 $\pm$    10 &     -5.3 &      464 $\pm$    27  \\
 4363 [O{\sc iii}]            &     0.149  &        59 $\pm$     3 &     -1.3 &       61 $\pm$     5   &        24 $\pm$     8 &     -0.2 &       25 $\pm$     8  \\
 4471 He{\sc i}               &     0.115  &        37 $\pm$     5 &     -0.7 &       38 $\pm$     6   &        36 $\pm$     6 &     -0.3 &       37 $\pm$     7  \\
 4686 He{\sc ii}              &     0.050  &        26 $\pm$     5 &     -0.5 &       26 $\pm$     5   &        $\cdots$       & $\cdots$ &         $\cdots$      \\
 4861 H$\beta$                &     0.000  &      1000 $\pm$     8 &    -27.1 &     1000 $\pm$    47   &      1000 $\pm$    10 &    -12.1 &     1000 $\pm$    47  \\
 4959 [O{\sc iii}]            &    -0.026  &      1170 $\pm$     9 &    -26.9 &     1163 $\pm$    53   &       607 $\pm$    11 &     -6.2 &      602 $\pm$    29  \\
 5007 [O{\sc iii}]            &    -0.038  &      3471 $\pm$    11 &    -75.0 &     3438 $\pm$   153   &      1750 $\pm$    12 &    -16.7 &     1732 $\pm$    78  \\
 5876 He{\sc i}               &    -0.203  &        96 $\pm$    14 &     -2.8 &       91 $\pm$    14   &       104 $\pm$    20 &     -1.3 &       98 $\pm$    19  \\
 6312 [S{\sc iii}]            &    -0.264  &        19 $\pm$     2 &     -0.6 &       18 $\pm$     2   &        22 $\pm$     6 &     -0.3 &       20 $\pm$     5  \\
 6548 [N{\sc ii}]             &    -0.296  &        51 $\pm$    11 &     -1.6 &       47 $\pm$    10   &        51 $\pm$    12 &     -0.8 &       48 $\pm$    11  \\
 6563 H$\alpha$               &    -0.298  &      3000 $\pm$    23 &    -92.8 &     2797 $\pm$    30   &      3026 $\pm$    16 &    -46.7 &     2814 $\pm$    21  \\
 6584 [N{\sc ii}]             &    -0.300  &       145 $\pm$    11 &     -4.5 &      135 $\pm$    10   &       149 $\pm$     6 &     -2.2 &      138 $\pm$     6  \\
 6678 He{\sc i}               &    -0.313  &        27 $\pm$     5 &     -0.8 &       25 $\pm$     4   &        $\cdots$       & $\cdots$ &         $\cdots$      \\
 6717 [S{\sc ii}]             &    -0.318  &       307 $\pm$     7 &     -9.4 &      285 $\pm$     7   &       300 $\pm$    17 &     -4.4 &      277 $\pm$    16  \\
 6731 [S{\sc ii}]             &    -0.320  &       236 $\pm$     8 &     -7.2 &      219 $\pm$     8   &       206 $\pm$    14 &     -3.0 &      191 $\pm$    13  \\
 7065 He{\sc i}               &    -0.364  &        23 $\pm$     5 &     -0.7 &       21 $\pm$     4   &        $\cdots$       & $\cdots$ &         $\cdots$      \\
 7136 [Ar{\sc iii}]           &    -0.374  &        87 $\pm$    11 &     -2.7 &       80 $\pm$    10   &        80 $\pm$    16 &     -1.3 &       73 $\pm$    15  \\
 7319 [O{\sc ii}]             &    -0.398  &        44 $\pm$     4 &     -1.4 &       40 $\pm$     4   &        $\cdots$       & $\cdots$ &         $\cdots$      \\
 7330 [O{\sc ii}]             &    -0.400  &        42 $\pm$     3 &     -1.3 &       38 $\pm$     2   &        $\cdots$       & $\cdots$ &         $\cdots$      \\
\hline
I(H$\beta$)(erg\,s$^{-1}$\,cm$^{-2}$) & & \multicolumn{3}{c}{  3.93e-15} &  \multicolumn{3}{c}{  2.61e-15} \\
c(H$\beta$) & & \multicolumn{3}{c}{0.11 $\pm$ 0.02} &  \multicolumn{3}{c}{0.12 $\pm$ 0.02} \\
A(V) & & \multicolumn{3}{c}{0.24 $\pm$ 0.04} &  \multicolumn{3}{c}{0.26 $\pm$ 0.04} \\
\hline
\end{tabular}
\end{table*}

\addtocounter{table}{-1}
\begin{table*}
 \caption{-- {\it Continued}}
\begin{tabular}{lc@{\hspace{3em}}ccc}
\hline
\multicolumn{1}{l}{} & \multicolumn{1}{l}{} & \multicolumn{3}{c}{\hspace{-3em}\large NGC 6789-E} \\
\multicolumn{1}{c}{$\lambda$ ({\AA})} & f($\lambda$) & F($\lambda$) & EW(\AA) & I($\lambda$) \\
\hline
 3726 [O{\sc ii}]            &    0.322 &      347 $\pm$    51 &    -30.5 &      492 $\pm$    94  \\
 3729 [O{\sc ii}]            &    0.322 &      398 $\pm$    25 &    -28.3 &      563 $\pm$    77  \\
 3868 [Ne{\sc iii}]          &    0.291 &       62 $\pm$    11 &     -2.2 &       86 $\pm$    18  \\
 4102 H$\delta$              &    0.229 &      145 $\pm$    10 &     -7.6 &      186 $\pm$    25  \\
 4340 H$\gamma$              &    0.157 &      374 $\pm$     9 &    -11.9 &      443 $\pm$    48  \\
 4363 [O{\sc iii}]           &    0.149 &       20 $\pm$     9 &     -0.6 &       23 $\pm$    11  \\
 4471 He{\sc i}              &    0.115 &       36 $\pm$     5 &     -1.7 &       41 $\pm$     7  \\
 4861 H$\beta$               &    0.000 &     1000 $\pm$    13 &    -35.7 &     1000 $\pm$    93  \\
 4959 [O{\sc iii}]           &   -0.026 &      842 $\pm$    15 &    -22.9 &      818 $\pm$    75  \\
 5007 [O{\sc iii}]           &   -0.038 &     2358 $\pm$    17 &    -60.0 &     2263 $\pm$   201  \\
 5876 He{\sc i}              &   -0.203 &      124 $\pm$    23 &     -4.3 &      100 $\pm$    20  \\
 6548 [N{\sc ii}]            &   -0.296 &       42 $\pm$    14 &     -1.3 &       39 $\pm$    13  \\
 6563 H$\alpha$              &   -0.298 &     3090 $\pm$    32 &    -98.9 &     2836 $\pm$    42  \\
 6584 [N{\sc ii}]            &   -0.300 &       69 $\pm$    12 &     -2.1 &       64 $\pm$    11  \\
 6717 [S{\sc ii}]            &   -0.318 &      225 $\pm$    23 &     -8.4 &      202 $\pm$    21  \\
 6731 [S{\sc ii}]            &   -0.320 &      207 $\pm$    18 &     -8.4 &      186 $\pm$    17  \\
 7136 [Ar{\sc iii}]          &   -0.374 &       80 $\pm$     8 &     -2.8 &       68 $\pm$     6  \\
 7319 [O{\sc ii}]            &   -0.398 &       27 $\pm$     4 &     -1.2 &       22 $\pm$     3  \\
 7330 [O{\sc ii}]            &   -0.400 &       20 $\pm$     8 &     -1.0 &       17 $\pm$     6  \\
\hline
I(H$\beta$)(erg\,s$^{-1}$\,cm$^{-2}$) & & \multicolumn{3}{c}{  2.94e-15}\\
c(H$\beta$) & & \multicolumn{3}{c}{0.47 $\pm$ 0.06}\\
A(V) & & \multicolumn{3}{c}{1.00 $\pm$ 0.10} \\
\hline
\end{tabular}
\end{table*}

\subsection{Electron densities and temperatures}
\label{elden}

\begin{table*}
\caption{Electron densities and temperatures for the five knots. Densities units are in 
cm$^{-3}$ and temperatures in 10$^{4}$ K.}
\label{temden}
\begin{tabular}{l  c  c  c  c  c }
\hline
                                &         NGC~6789-A         &         NGC~6789-B         &         NGC~6789-C         &         NGC~6789-D         &        NGC~6789-E          \\
\hline
n([S{\sc ii}])                  &        80:                 &        60:                 &        60:                 &       90:                  &       300:                 \\
n([O{\sc ii}])                  &        60:                 &        70:                 &        60:                 &       70:                  &       250:                 \\
t([O{\sc iii}])                 &      1.44  $\pm$       0.05&      1.36  $\pm$       0.10&      1.44  $\pm$       0.05&      1.30  $\pm$       0.15&      1.16  $\pm$       0.18\\
t([O{\sc ii}])                  &      1.35  $\pm$ 0.06$^{a}$&      1.18  $\pm$       0.10&      1.41  $\pm$       0.08&      1.21  $\pm$ 0.10$^{a}$&      1.33  $\pm$       0.25\\
t([S{\sc iii}])$^{b}$           &      1.38  $\pm$       0.16&      1.30  $\pm$       0.19&      1.38  $\pm$       0.16&      1.22  $\pm$       0.24&      1.06  $\pm$       0.25\\
\hline
\end{tabular}
\begin{flushleft}
$^{a}$ From a relation with T([O{\sc iii}]) based on photoionisation models\\
$^{b}$ From an empirical relation with T([O{\sc iii}])
\end{flushleft}
\end{table*}

The electron density and temperatures of the ionised gas were derived from the emission line 
data using the same procedures as in \cite{pmontero03}, based on the five-level statistical 
equilibrium atom approximation in the IRAF task \texttt{temden} \citep{derobertis87,shaw95}. 
See \cite{hagele08} and Appendix B of \cite{garcia-benito09} for a description of the 
temperature and densities relations, respectively. We took as sources of error the
uncertainties associated with the reddening-corrected emission-line fluxes and
we propagated them through our calculations. 

Electron densities were derived from the [S{\sc ii}] $\lambda\lambda$ 6717,6731 \AA\ and 
[O{\sc ii}] $\lambda\lambda$ 3726,3729 \AA\ line ratios, which are representative of the 
low-excitation zone of the ionised gas. In all cases they provide upper limits, which are 
remarkably similar for both ions. In the case of the oxygen lines, the spectral dispersion 
does not allow the total resolution of the lines, and they were deblended by a multi-Gaussian fit. 
The upper limits for the electron density are lower in all cases than the critical value for 
collisional de-excitation.

We computed three electron temperatures, T([O{\sc ii}]), T([O{\sc iii}]) and T([S{\sc iii}]), for 
each of the five knots. The auroral [O{\sc iii}] $4363$ was detected in all the five knots with 
sufficient signal-to-noise, and the [O{\sc iii}] electron temperature was derived directly from 
the ratio (I(4959 \AA) + I(5007 \AA))/ I(4363 \AA).

The auroral [O{\sc ii}] $7319$ and [O{\sc ii}] $7330$ lines from the 2F 
multiplet are emission-line doublets. Due to the spectral resolution, we report here the sum of 
each doublet. Using the calculated [O{\sc iii}] electron temperatures, we checked that the 
contribution by direct recombination is negligible for these lines \citep{liu00}. For knot A and 
D, the intensity of the [O{\sc ii}] $\lambda\lambda$ 7319,7330 \AA\ did not allow an accurate 
measure, therefore we derived their [O{\sc ii}] temperature from T([O{\sc iii}]) using the relation 
based on the photoionisation models described in \cite{pmontero03}, which takes into account 
explicitly the dependence of T([O{\sc ii}]) on the electron density.

Although we could measure the [S{\sc iii}] $\lambda$ 6312 \AA\ in some of the knots, we were not 
able to get the [S{\sc iii}] $\lambda$ 9069,9532 \AA\ lines, so we estimated the [S{\sc iii}] 
temperature from the empirical relation

\begin{equation}
t([\textrm{{S\sc iii}}]) = (1.19 \pm 0.08) t([\textrm{{O\sc iii}}]) - (0.32 \pm 0.10)
\end{equation}

\noindent found by \cite{hagele06}.

The electron densities and temperatures derived in the five knots of NGC~6789 are listed in 
Table \ref{temden} along with their corresponding errors.

\begin{table*}
\caption{Ionic chemical abundances for helium.}
\label{helium}
\begin{tabular}{l  c  c  c  c  c }
\hline
                                &              NGC 6789-A    &             NGC 6789-B     &         NGC 6789-C         &         NGC 6789-D         &          NGC 6789-E       \\
\hline
He$^{+}$/H$^{+}$ (4471)         &     0.077  $\pm$      0.011&     0.080  $\pm$      0.015&     0.079  $\pm$      0.013&     0.076  $\pm$      0.014&     0.082  $\pm$      0.015 \\
He$^{+}$/H$^{+}$ (5876)         &     0.075  $\pm$      0.011&     0.071  $\pm$      0.013&     0.072  $\pm$      0.012&     0.076  $\pm$      0.015&     0.073  $\pm$      0.016\\
He$^{+}$/H$^{+}$ (6678)         &   $\cdots$                 &           $\cdots$         &     0.071  $\pm$      0.013&      $\cdots$              &   $\cdots$ \\
He$^{+}$/H$^{+}$ (7065)         &   $\cdots$                 &   $\cdots$                 &     0.074  $\pm$      0.016&      $\cdots$              &   $\cdots$ \\
He$^{+}$/H$^{+}$                &     0.076  $\pm$      0.010&     0.075  $\pm$      0.012&     0.074  $\pm$      0.009&     0.076  $\pm$      0.011&     0.077  $\pm$      0.011\\
He$^{2+}$/H$^{+}$ (4686)        &   $\cdots$                 &   $\cdots$                 &    0.0023  $\pm$     0.0006&     $\cdots$               &   $\cdots$ \\
He/H                            &   $\cdots$                 &   $\cdots$                 &     0.076  $\pm$      0.009&      $\cdots$              &   $\cdots$ \\
\hline
\end{tabular} 
\end{table*}

\subsection{Ionic abundances}
\label{abundances}

We derived the ionic abundances of the different chemical species using the brightest available 
emission lines detected in the optical spectra using the task \texttt{ionic} of the STSDAS package 
in {\sc iraf} \citep[see][]{hagele08}. The total abundances were derived by taking into account, 
when required, the unseen ionisation stages of each element, resorting to the ionisation correction 
factors (ICFs) for each species derived from the photoionisation models described in Section 
\ref{stion} below.

\begin{equation}
\frac{X}{H} = ICF(X^{i})\frac{X^{+i}}{H^{+}}
\end{equation}

\subsubsection{Helium abundance}

Four of the strongest helium emission lines, He{\sc i} $\lambda\lambda$ 4471, 5876, 6678, and 7065  
\AA\ were used to calculate He$^{}+$/H$^{}+$. The last two lines were only detected with enough 
signal-to-noise in knot C. Also in this knot, He {\sc ii} $\lambda$ 4686 was measured, allowing the 
calculation of twice ionised He. 

Helium lines arise mainly from pure recombination; however, they could have some contribution from 
collisional excitation as well as being affected by self-absorption and, if present, by underlying 
stellar absorptions \citep[see][for a complete treatment]{olive01}, but we checked
that these absorptions produce negligible variations in the measure of the
lines in the STARLIGHT subtracted spectra.

We took the electron temperature of [O{\sc iii}] as representative of the zone where the He 
emission arises and we used the equations given by \cite{olive01} to derive the He$^{+}$/H$^{+}$, 
using the theoretical emissivities scaled to H$\beta$ from \cite{smits96} and the expressions 
for the collisional correction factors from \cite{kingdon95}. To calculate the helium twice 
ionised, we used the equation found by \cite{kunth83}. We did not take into account, however, 
the corrections for fluorescence since the involved helium lines have negligible dependence 
on optical-depth effects and the observed knots have low densities. 

The results obtained for each line and their corresponding errors are presented in Table 
\ref{helium}. For knot C, the total abundance of He was calculated by adding directly the 
two ionic abundances, He/H = (He$^{+}$ + He$^{2+}$)/H$^{+}$. Also given in the table is 
the adopted value for He$^{+}$/H$^{+}$, a weighted average of the values, using the error 
of each line as weight.

\subsubsection{Ionic abundances from forbidden lines}

The oxygen ionic abundance ratios, O$^{+}$/H$^{+}$ and O$^{2+}$/H$^{+}$, were derived from
the [O{\sc ii}]\, $\lambda\lambda$\,3726,3729\,\AA\ and
[O{\sc iii}] $\lambda\lambda$\,4959, 5007\,\AA\ lines, respectively using the appropriate
electron temperature for each ion.

The ionic abundance of nitrogen, N$^{+}$/H$^{+}$, was derived from the intensities of the 
[N{\sc ii}]\, $\lambda\lambda$\,6548,6584\,\AA\ lines assuming that T([N{\sc ii}]) $\approx$ 
T([O{\sc ii}]). 

For sulphur, we derived S$^{+}$/H$^{+}$ abundances from the fluxes of the [S{\sc ii}] emission lines 
at $\lambda\lambda$ 6717,6731 \AA\ assuming that T([S{\sc ii}]) $\approx$ T([O{\sc ii}]). 
Since we were not able to measure the near infrared [S{\sc iii}] $\lambda\lambda$ 9069,9532 \AA\, 
we derived the S$^{2+}$/H$^{+}$ abundances using T([S{\sc iii}]) and [S{\sc iii}] $\lambda$ 6312 \AA\ 
by means of the equation
 
\begin{align}
12\,+\,log(S^{2+}/H^{+})\,&=&\,log\frac{I(6312)}{I(H\beta)}\,+\,6.74 \nonumber \\
& & +\,\frac{1.672}{t}\,-\,0.595\,log\,t
\end{align}

Neon ionic abundance was estimated from the [Ne {\sc iii}] emission line at $\lambda$ 3869 \AA. 
For this ion, we took the electron temperature of [O{\sc iii}] as representative of the high 
excitation zone \citep{peimbert69}. 

[Ar {\sc iii}] $\lambda$ 7136 \AA\ was the only argon line detected in the spectra and  
the abundance of Ar$^{2+}$ was calculated assuming that T([Ar{\sc iii}]) $\approx$ 
T([S{\sc iii}]) \citep{garnett92}. 

The ionic abundances -and their corresponding errors- for each observed element for 
the five knots are given in Table \ref{ionic}.

\begin{table*}
\caption{Ionic chemical abundances derived from forbidden emission lines, ICFs$^{a}$ and total
chemical abundances for elements heavier than helium.}
\label{ionic}
\begin{tabular}{l  c  c  c  c  c }
\hline
 &        NGC~6789-A &        NGC~6789-B &        NGC~6789-C &        NGC~6789-D &        NGC~6789-E \\
\hline
12 + log(O$^{+}$/H$^{+}$)       &      7.43  $\pm$       0.04&      7.82  $\pm$       0.11&      7.42  $\pm$       0.07&      7.73  $\pm$       0.11&      7.16  $\pm$       0.23\\      
12 + log(O$^{2+}$/H$^{+}$)      &      7.63  $\pm$       0.04&      7.28  $\pm$       0.07&      7.59  $\pm$       0.04&      7.41  $\pm$       0.13&      7.68  $\pm$       0.17\\
\textbf{12 + log(O/H)}          &      7.84  $\pm$       0.04&      7.93  $\pm$       0.10&      7.81  $\pm$       0.05&      7.90  $\pm$       0.12&      7.80  $\pm$       0.18\\
12 + log(N$^{+}$/H$^{+}$)       &      6.03  $\pm$       0.04&      6.37  $\pm$       0.07&      6.08  $\pm$       0.06&      6.22  $\pm$       0.07&      5.85  $\pm$       0.16\\
ICF(N$^{+}$)                    &      3.57                  &      1.57                  &      4.57                  &      1.97                  &      4.09                  \\
\textbf{12 + log(N/H)}          &      6.58  $\pm$       0.04&      6.57  $\pm$       0.07&      6.74  $\pm$       0.06&      6.51  $\pm$       0.07&      6.50  $\pm$       0.16\\
log(N/O)                        &     -1.26  $\pm$       0.06&     -1.37  $\pm$       0.12&     -1.07  $\pm$       0.08&     -1.38  $\pm$       0.14&     -1.30  $\pm$       0.24\\
12 + log(S$^{+}$/H$^{+}$)       &      5.70  $\pm$       0.03&      5.91  $\pm$       0.07&      5.75  $\pm$       0.05&      5.83  $\pm$       0.07&      5.70  $\pm$       0.14\\
12 + log(S$^{2+}$/H$^{+}$)      &      5.88  $\pm$       0.16&      6.14  $\pm$       0.23&      6.09  $\pm$       0.16&      6.35  $\pm$       0.30&         $\cdots$           \\
ICF(S$^{+}$ + S$^{2+}$)         &      1.11                  &      1.01                  &      1.24                  &      1.01                  &      7.64$^{b}$           \\
\textbf{12 + log(S/H)}          &      6.20  $\pm$       0.11&      6.35  $\pm$       0.18&      6.41  $\pm$       0.11&      6.48  $\pm$       0.25&      6.58 $\pm$        0.14\\
log(S/O)                        &     -1.64  $\pm$       0.12&     -1.58  $\pm$       0.20&     -1.41  $\pm$       0.12&     -1.41  $\pm$       0.27&     -1.22 $\pm$       -0.25\\
12 + log(Ne$^{2+}$/H$^{+}$)     &      7.00  $\pm$       0.06&      6.77  $\pm$       0.10&      7.00  $\pm$       0.06&      6.68  $\pm$       0.17&      6.75  $\pm$       0.20\\
ICF(Ne$^{2+}$)                  &      1.39                  &      1.61                  &      1.28                  &      2.07                  &      1.22                   \\
\textbf{12 + log(Ne/H)}         &      7.14  $\pm$       0.06&      6.98  $\pm$       0.10&      7.11  $\pm$       0.06&      7.00  $\pm$       0.17&      6.82  $\pm$       0.20\\
log(Ne/O)                       &     -0.70  $\pm$       0.07&     -0.95  $\pm$       0.14&     -0.70  $\pm$       0.08&     -0.90  $\pm$       0.21&     -0.96  $\pm$       0.27\\
12 + log(Ar$^{2+}$/H$^{+}$)     &      5.67  $\pm$       0.08&      5.56  $\pm$       0.12&      5.56  $\pm$       0.09&      5.63  $\pm$       0.17&      5.70  $\pm$       0.19\\
ICF(Ar$^{2+}$)                  &      1.12                  &      1.21                  &      1.15                  &      1.14                  &      1.10                  \\
\textbf{12 + log(Ar/H)}         &      5.72  $\pm$       0.08&      5.64  $\pm$       0.12&      5.62  $\pm$       0.09&      5.68  $\pm$       0.17&      5.74  $\pm$       0.19\\
log(Ar/O)                       &     -2.12  $\pm$       0.09&     -2.28  $\pm$       0.16&     -2.19  $\pm$       0.10&     -2.22  $\pm$       0.21&     -2.06  $\pm$       0.22\\
\hline
\end{tabular}
\begin{flushleft}
$^{a}$ ICFs estimated from tailored photoionisation models (see Section \ref{stion})\\
$^{b}$ ICF(S$^+$)
\end{flushleft}
\end{table*}

\section{Optical and UV photometry}
\label{phot}

\subsection{H$\alpha$ and GALEX photometry}
\label{Halfa}

We analysed the H$\alpha$ image of the galaxy, retrieved from the Palomar/Las Campanas atlas 
of blue compact dwarf galaxies \citep{gildepaz03}. We defined elliptical apertures on this image 
for each of the five knots extracted in the spectroscopic observations and we measured all the 
flux inside the elliptical apertures up to the isophote corresponding to the 50\% of the peak 
of the intensity for each knot (see Figure \ref{regslit}). The total fluxes and the 
corresponding luminosities at the adopted distance are listed in Table \ref{hafot} with their 
corresponding errors. The observed H$\alpha$ fluxes were corrected in each knot for dust extinction 
using the values of c(H$\beta$), given in Table \ref{lines}. We also list in the same table the size 
of each knot, as the radius of the circular aperture whose area is equal to that encompassed by the 
elliptical aperture. As can be seen in Figure \ref{regslit}, NGC 6789 has an H$\alpha$ major 
axis of 740 pc and 560 pc for its minor axis, measured using an elliptical aperture that 
encompasses the isophote at 3$\sigma$ over the level of the sky background noise. The total flux 
inside this isophote (not extinction-corrected) 
is (2.0 $\pm$ 0.1) $\times$ 10$^{-12}$ erg s$^{-1}$ cm$^{-2}$. 

\begin{table*}
\caption{Properties of the individual knots as measured in the H$\alpha$ and GALEX photometry.}
\label{hafot}
\begin{tabular}{l  c  c  c  c  c }
\hline
                                &              NGC 6789-A    &             NGC 6789-B     &         NGC 6789-C         &         NGC 6789-D         &          NGC 6789-E       \\
\hline
log F(H$\alpha$) (erg/s/cm$^2$)   &  -13.96 $\pm$ 0.03 & -13.85 $\pm$  0.03 & -13.69 $\pm$ 0.03 & -14.03 $\pm$ 0.03 & -14.17  $\pm$  0.03 \\
log L(H$\alpha$) (erg/s)         &  37.23  $\pm$ 0.02 &  37.34 $\pm$ 0.02 &  37.50 $\pm$ 0.03 & 37.17 $\pm$ 0.02 & 37.02 $\pm$      0.01\\
Radius (pc)         &   30     &   34   &    37 &   36    &    21 \\
FUV (mag) &   20.00 $\pm$ 0.14 & 20.00 $\pm$ 0.16 & 20.18 $\pm$ 0.20 & 19.71 $\pm$ 0.15 & 21.24 $\pm$ 0.04 \\
NUV (mag) &   19.86 $\pm$ 0.08 & 19.69 $\pm$ 0.09 & 19.86 $\pm$ 0.10 & 19.68 $\pm$ 0.09 & 21.18 $\pm$ 0.01 \\
FUV - NUV (mag)         &   0.14 $\pm$ 0.16  &   0.28 $\pm$ 0.19  &     0.31 $\pm$  0.22 & 0.03 $\pm$ 0.18  &  0.06 $\pm$ 0.04 \\
\hline
\end{tabular} 
\end{table*}

The \textit{Galaxy Evolution Explorer} \citep[GALEX;][]{martin05,morrissey05} far-ultraviolet 
(FUV; $\lambda_{ref}$ = 1530 \AA, $\Delta\lambda$ = 400 \AA) and near-ultraviolet 
(NUV; $\lambda_{ref}$ = 2310 \AA, $\Delta\lambda$ = 1000 \AA) images of NGC 6789 were retrieved 
from the Nearby Galaxies Survey (NGS). We used the same elliptical apertures defined in the 
H$\alpha$ image to obtain the FUV and NUV fluxes in each knot. In Table \ref{hafot} we show for each 
knot both the FUV and NUV magnitudes once extinction corrected together with the colour index 
FUV-NUV, all of them in AB mags. 
As typical sizes of the studied star-forming knots are lower than the GALEX point-spread function 
(PSF), some aperture effects to the measured UV fluxes that cannot be quantified may exist.
However, the characterization of the knots by the corresponding colour indices are
less affected by this effect than the total flux measurements in these regions,
as we checked by taking different aperture sizes around the position of
the knots.

As can be seen, the four brightest knots (from A to D) have similar
UV luminosities both in FUV and NUV, being the knot D the brightest one, although this is not
the brightest knot in H$\alpha$. Regarding colours, knots B and C present redder UV colours
as compared with the other three knots.

\subsection{WFPC2 Photometry \label{rphot}}

\begin{figure*}
\begin{minipage}{\textwidth}
\centering
\includegraphics[width=0.49\textwidth,clip=]{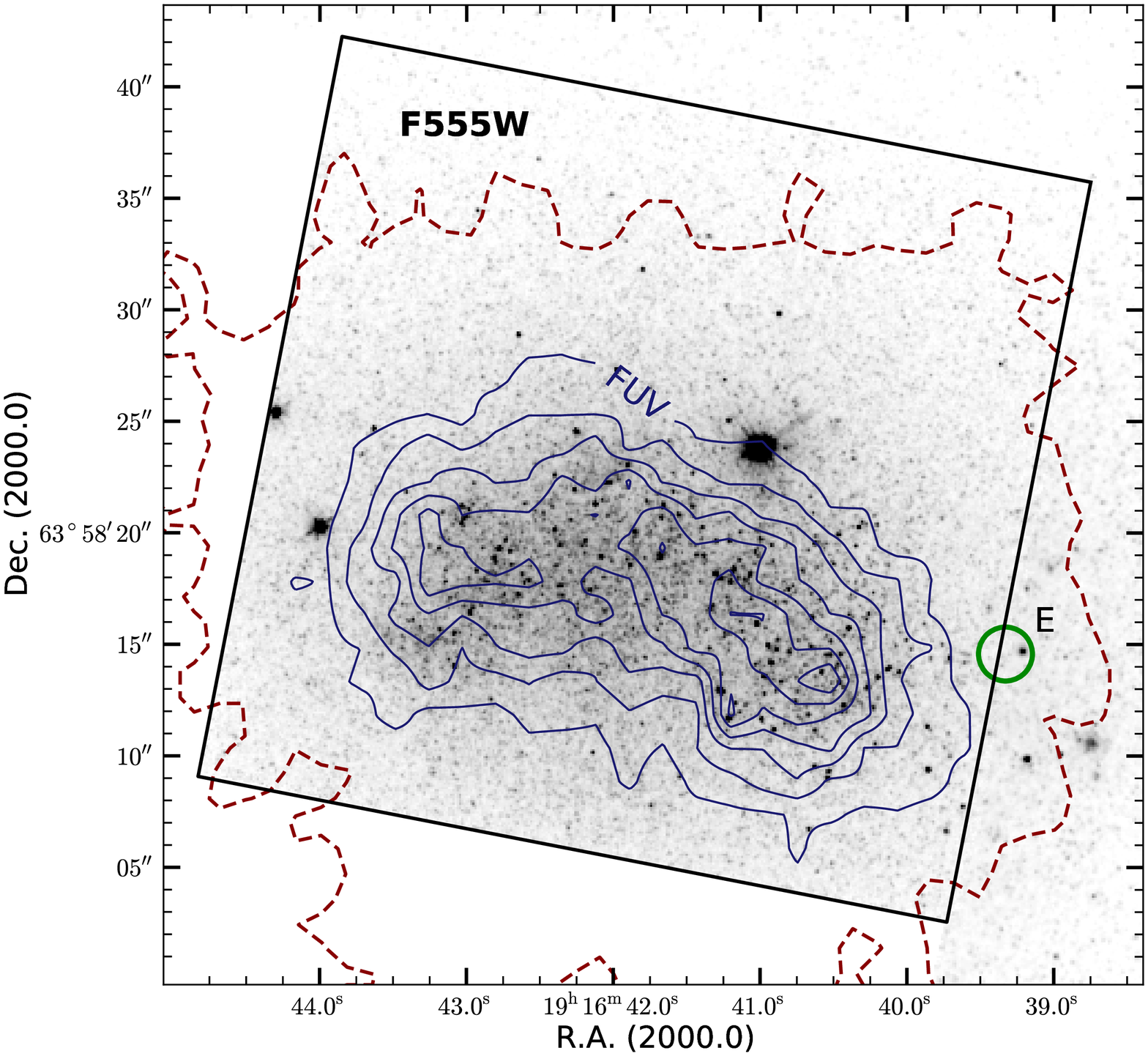}
\includegraphics[width=0.49\textwidth,clip=]{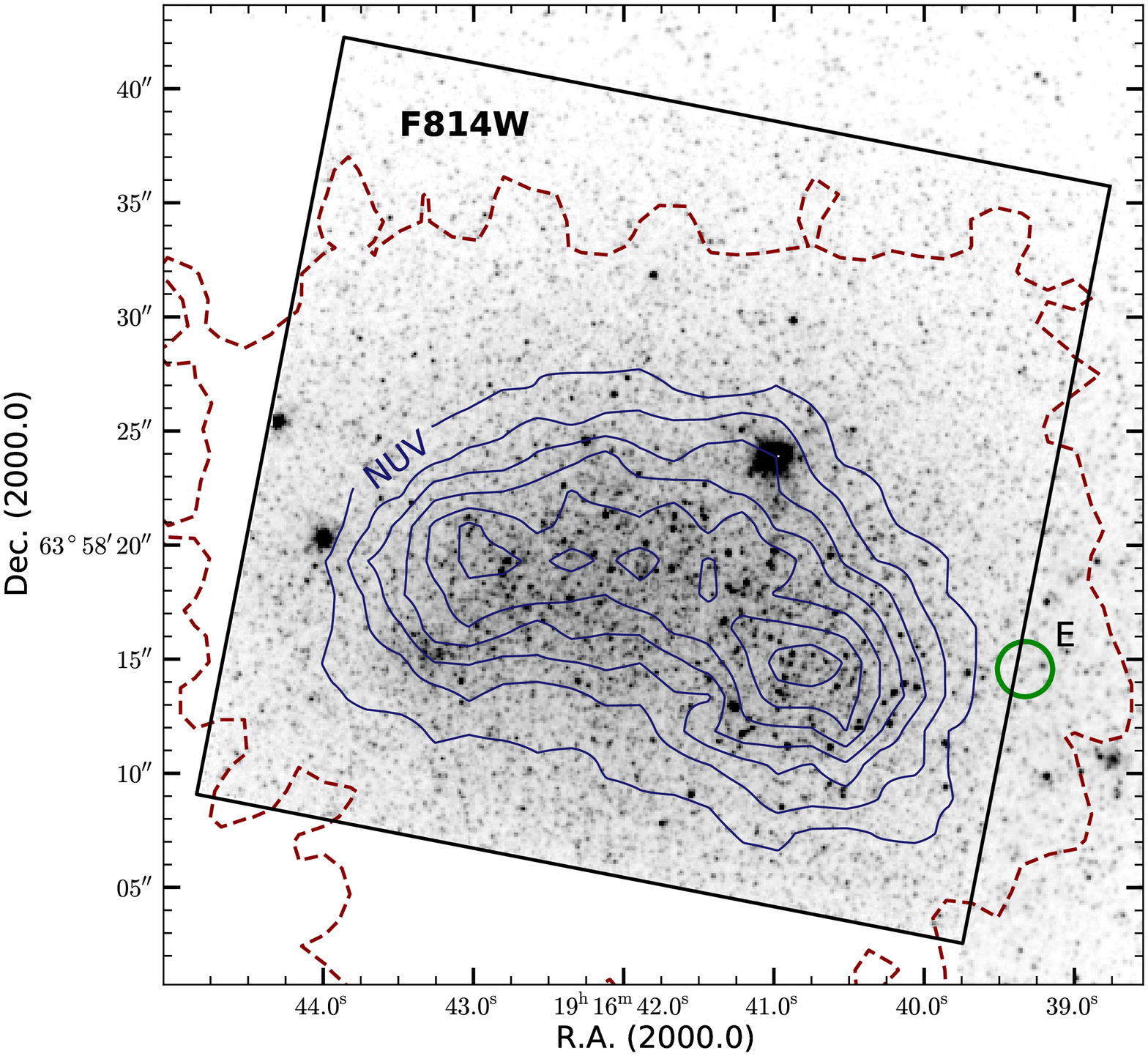}
\end{minipage}
\caption{HST WFPC2 images of NGC 6789 in continuum filters F555W and F814W. The solid contours 
are drawn from FUV and NUV GALEX images, respectively. The dashed contour corresponds to the 
isophote at 3$\sigma$ over the level of the sky background of the Palomar/Las Campanas H$\alpha$ 
image. The aperture of knot E defined in the H$\alpha$ image and the rectangle encompassing the 
FOV of the PC chip of the WFPC2 data are also shown for reference. North is up and east is towards 
the left-hand side.}
\label{hstgalex}
\end{figure*}

\textit{Hubble Space Telescope} (HST) Wide Field Planetary Camera 2 (WFPC2)
images of NGC~6789 were retrieved from the HST archive. We discuss photometry 
based on the images taken in 2000 July-September (GO-8122) through two continuum 
filters: F555W (V) and F814W (I). Each camera images onto a
Loral 800$\times$800 CCD which gives a plate scale of 0\farcs046 pixel$^{-1}$
for the PC camera and 0\farcs10 pixel$^{-1}$ for the three WF cameras, with a
readout noise of $\sim$5 e$^{-}$ and a gain of 7 e$^{-}/DN$ for this observations.
The central star-forming region of NGC~6789 was centered in the PC camera in all 
images using three different pointings. The scale of the PC CCD at NGC~6789, for which we 
assumed a distance modulus of $(m - M) = 27.80$ \citep{drozdovsky01}, is $0.80$ pc pixel$^{-1}$. 
Table \ref{log} lists the details concerning the WFPC2 data. Figure \ref{hstgalex} 
shows the HST WFPC2 images of NGC 6789 in F555W and F814W filters. The contours are 
drawn from FUV and NUV GALEX images, respectively. The aperture of knot E defined in the 
H$\alpha$ image and the rectangle encompassing the FOV of the PC chip of the WFPC2 data are 
also shown for reference.

\begin{table*}
\caption{Journal of HST/WFPC2 observations of NGC~6789. The images were 
obtained in 2000 July-September for the cycle 8 program GO-8122, with 
Regina Schulte-Ladbeck as PI. The object of study is centered in the PC chip.}
\label{log}
\begin{tabular}{l  c  c  c  c  c }
\hline
Filter & RA & Dec & Exposure  & Observation ID\\
       & (J2000) & (J2000) & (s)  & \\
\hline
F555W & 19:16:37.77 & +63:58:37.2 & 2 $\times$ 1300 &  U5BF0301R, U5BF0302R \\
      & 19:16:37.73 & +63:58:37.4 & 2 $\times$ 1400 &  U5BF0303R, U5BF0304R \\
      & 19:16:37.68 & +63:58:37.6 & 2 $\times$ 1400 &  U5BF0305R, U5BF0306R \\
F814W & 19:16:37.77 & +63:58:37.2 & 2 $\times$ 1300 &  U5BF0401R, U5BF0402R \\
      & 19:16:37.73 & +63:58:37.4 & 2 $\times$ 1400 &  U5BF0403R, U5BF0404R \\
      & 19:16:37.68 & +63:58:37.6 & 2 $\times$ 1400 &  U5BF0405R, U5BF0406R \\ 
\hline
\end{tabular}
\end{table*}

The stellar photometric analysis was performed with the HSTphot package \citep{dolphin00a}. 
This package is specifically designed for use with HST WFPC2 images and uses a library of 
Tiny Tim \citep{krist95} undersampled PSFs for different locations 
of the star on the camera and of the star within the pixel, to centre the star and 
to find its magnitude, given in the flight system magnitude. The first step was to run 
the \texttt{mask} task using the HST data quality file (\texttt{c1f}) to mask out the bad 
pixels and other image defects. The next step was to run \texttt{crmask} for cosmic ray 
removal. We used a registration factor of 0.5 and $\sigma$-threshold of 3. It has the 
capability of cleaning images that are not perfectly aligned, and it can handle images 
from different filters. After cosmic-ray rejection, sets of images of each filter at a common 
pointing were combined into a single image, using the routine \texttt{coadd}. Since 
there are two images per pointing, we ended up with 3 images per filter. The sky 
computation is made by \texttt{getsky}, which takes all pixels in an annulus around each
pixel, determines the sky value and calculates the sky background map. These sky values 
are used only as a starting guess in the HSTphot photometry. The final step requires the 
use of the \texttt{hotpixels} procedure on each combined image, which uses the result from 
\texttt{getsky} and tries to locate and remove all hot pixels. This is an important step, 
since hot pixels can create false detections and also, can throw off the PSF solutions.

The main \texttt{hstphot} routine was run on the images in the F555W and F814W bands. 
This task performs stellar PSF photometry on multiple images from different 
filters and pointings (providing the dithering pattern, it analyses the dithering image 
as a whole), including alignment and aperture corrections, as well as PSF modifications 
to correct for errors of geometric distortion via the \citet{holtzman95} distortion 
correction equations and the 34th row error, noted by \citet{shaklan95} (see also 
\citealt{anderson99}), and correction for charge transfer inefficiency \citep{dolphin00b}. 
We enabled the determination of a ``local sky'' value (option 2).

We selected ``good stars'' from the \texttt{hstphot} output. Object types were classified 
as good star, possible unresolved binary, bad star, single-pixel cosmic ray or hot pixel, 
and extended object. To ensure selecting high fidelity point sources, we also used 
the ``sharpness'' parameter (absolute value to be $\leq$ 0.35), a measure of the quality 
of the fit ($\chi^{2}$ $\leq$ 2.5) and a minimum signal-to-noise ratio of 10 to reject 
false star detections in regions with structured nebulosity or artifacts. The final number 
of stars detected in the PC chip with these parameters in both filters were 7410.

To quantify the completeness and systematic uncertainty of the photometry, a grid of 
artificial stars was generated on a 2-dimensional CMD and distributed according to the 
flux of the images with an artificial star routine provided by HSTphot. 
The parameters of the routine are the minimum and maximum of the measured colour and 
magnitude. The magnitude steps used were multiple of 0.5, while colour steps are by 0.25.
The artificial stars were distributed on the CMD in accordance with the number observed.
Approximately 60,000 fake stars were added in each image (at different trials, in order
to leave the crowding conditions unaltered) and were given random magnitudes and colours
in the observed range. The 50\% completeness of the F555W filter is reached at 26.9 
magnitude, while for the F814W filter is 25.6 mag.

\begin{figure}
\includegraphics[width=0.5\textwidth,clip=]{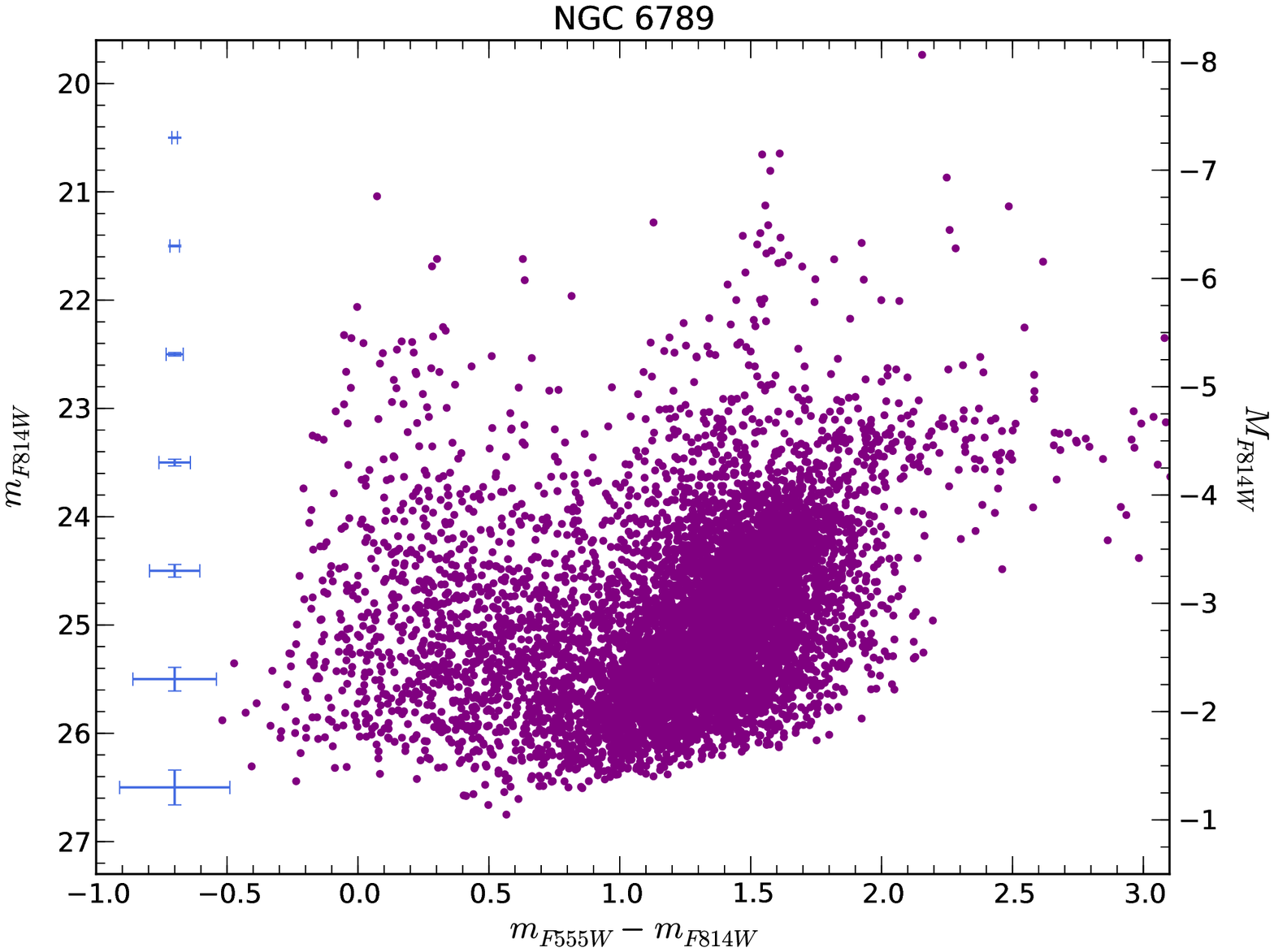}
\caption{CMD of NGC~6789 with average photometric uncertainties per 
magnitude bin.}
\label{cmd}
\end{figure}

The final CMD of NGC~6789 with typical photometric errors per magnitude bin is shown in 
Figure \ref{cmd}.

\section{Discussion}
\label{discussion}

\subsection{Star formation history and stellar populations \label{ssfh}}
Studies to obtain information on the star formation of composite stellar systems 
have been proven to be very successful (see e.g., \citealt{aparicio97}; 
\citealt{hernandez00}; \citealt{dolphin02}). Any SFH recovery method relies on 
the assumption that a composite stellar population can be considered simply as the 
combination of SSPs, assigning a certain relative weight to each SSP. The observed 
CMD is compared with theoretical ones created via Monte-Carlo methods for a variety 
of IMFs, binary fractions, star-formation laws, etc., extracting the stellar information 
from isochrones or stellar evolution tracks. Observed and theoretical CMDs are divided 
into boxes and converted into two-dimensional histograms of stellar density as function 
of colour and magnitude (Hess diagrams) and compared using statistical methods. 

For the analysis of this work, we used the StarFISH code\footnote{Available at 
\texttt{http://www.noao.edu/staff/jharris/SFH/}} developed by \cite{harris01}. This code 
was successfully used in a number of cases (e.g. \citealt{harris04}; 
\citealt{brown06}; \citealt{williams07}; \citealt{harris09}; \citealt{garcia-benito11}).
Using determinations of the interstellar extinction, photometric errors, and distance 
moduli, it uses minimisation of a chi-squared-like statistics technique to find the linear 
combination of single-component stellar population models that best fit an observed CMD.
The ages and metallicities of the underlying stellar population are characterized by 
the ages and metallicities of the CMDs included in the best, while SFR at each age is 
provided by the weights given to the CMDs.

The theoretical isochrones chosen for analysis were those of \cite{marigo08}\footnote{Available at 
\texttt{http://stev.oapd.inaf.it/cgi-bin/cmd/}} for ages in the range 1 Myr-14Gyr. This data set 
includes models with age bins spaced logarithmically since the CMD changes much more rapidly at 
young ages than at old ones. We chose the two metallicity values closest to the observed ones (see 
Section \ref{abundances}), namely Z = \{0.001, 0.004\}. \cite{marigo08} provide metallicities  
in the range 0.0001 $\leq$ Z $\leq$ 0.03. Nevertheless, for the sake of consistency we selected 
only the same values that were available for the STARLIGHT libraries \citep{bruzual03}. The 
photometric error and completeness estimates were taken directly from the results of the 
artificial star experiments described in Section \ref{rphot}. We adopted a Salpeter IMF with a 
spectral index of -1.35 from 0.1-100 M$_{\odot}$ \cite{salpeter55}. This assumption is likely 
to be valid since our CMD does not contain stars with masses $<$ 1 M$_{\odot}$. The binary 
fraction was set to a value of 0.5. 

Our reference value for NGC 6789 distance modulus is (m - M) = 27.80 $\pm$ 0.13 $\pm$ 0.18, value 
reported by \cite{drozdovsky01} using the tip of the red giant branch distance method. The Galactic 
extinction is A$_{V}$ = 0.212, value provided by \cite{schlegel98} extinction maps for an area 
with radius equal to 5 arcmin around NGC 6789, remarkably close to the reddening value obtained 
from our spectroscopic data (Section \ref{line}). Nevertheless, we built a set of models to 
explore the space of parameters. The SFH recovery is repeated for each point in the grid 
(m - M) vs A$_{V}$, and then we built a final $\chi^{2}_{min}$ map for the solutions. 
We explored the space of parameters in steps of 0.01 in distance modulus and extinction. To 
evaluate the errors of the recovered solutions, we generated a series of synthetic CMDs 
using the best-fitting SFR and find a correspondence between the $\chi^{2}_{min}$ 
and the confidence level of significance. 

Although it is tempting to perform a SFH analysis for each knot, the resulting individual CMDs, 
taken to be as the stars included inside the defined ellipses (see Figure \ref{regslit} 
and subsection \ref{Halfa}), do not contain enough stars\footnote{Specially Knot E, part of 
which falls outside the PC FOV.} and, therefore, the large errors associated with each individual 
SFH prevent us to draw any conclusion. At any rate, the distribution of stars in each individual 
CMD is very similar for all the knots. Thus, we use the entire CMD of the galaxy to derive the 
global SFH.

\begin{figure}
\includegraphics[width=0.5\textwidth,clip=]{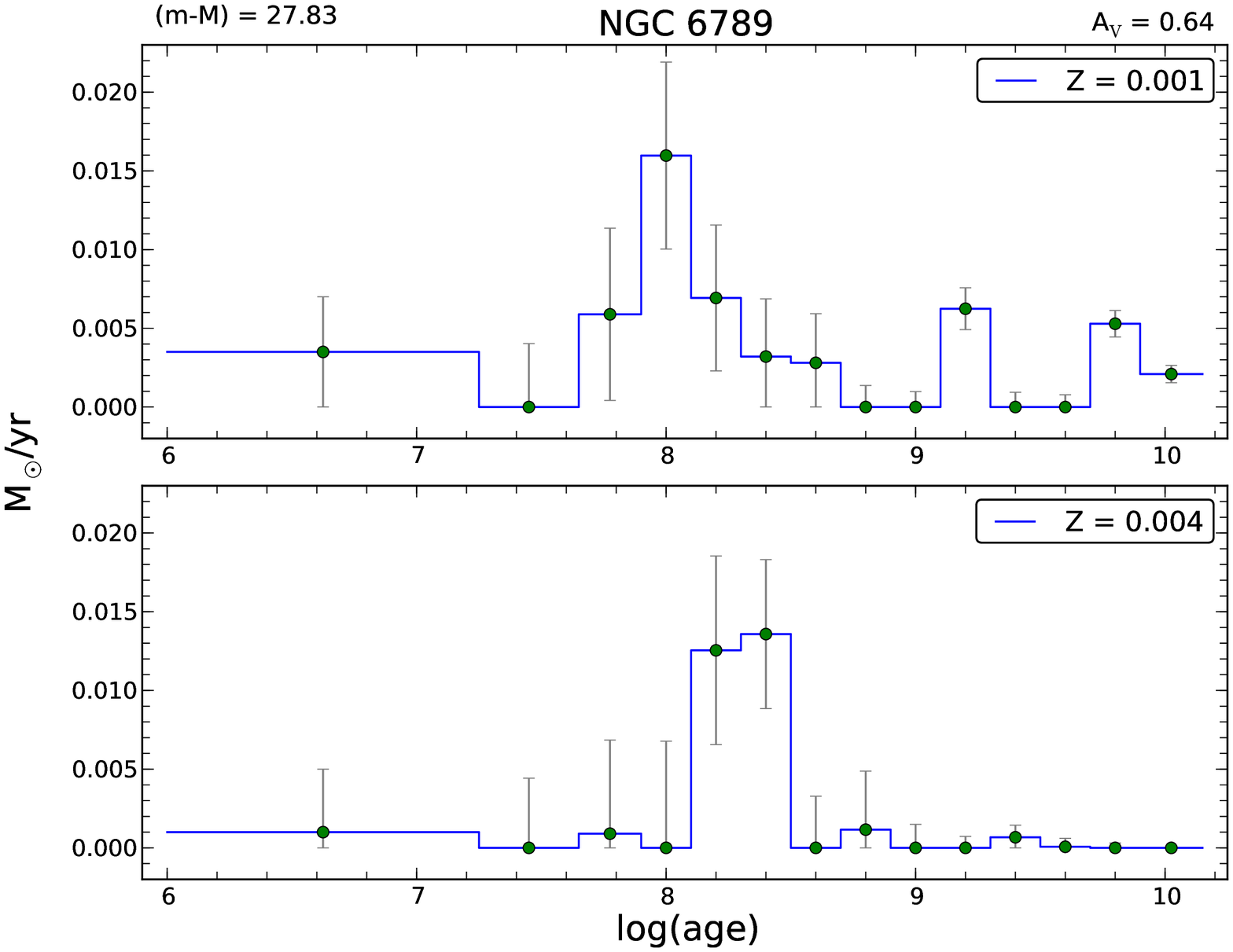}
\caption{Best StarFISH SFH fit derived from the HST optical observations of the resolved 
stellar populations for NGC 6789. The best distance modulus and extinction values are also 
shown.}
\label{sfh}
\end{figure}

Figure \ref{sfh} shows the overall best-fitting StarFISH solution for the SFH of NGC 6789, located 
at (m - M) = 27.83 $\pm$ 0.06 and A$_{V}$ = 0.64 $\pm$ 0.08, where the errors stand for the 
1$\sigma$ confidence level. The errors in the star formation diagram only allow the consideration 
of a few main bursts in the SFH. Regarding the lower metallicity (Z = 0.001), 
the solution shows a few bursty events around 1.6 Gyr and 6-10 Gyr, followed by a more recent 
burst around 100 Myr and a very young one during the last 12 Myr. As for the higher metallicity 
(Z = 0.004), only one event between 150-300 Myr is clearly seen. Although for the very young 
stars ($<$ 12 Myr) both metallicities present a significant error, we take Z = 0.001 as the 
metallicity for the youngest event, as derived from the nebular analysis (see section 
\ref{abundances}). The SFR, according to StarFISH's results, was stronger around 100 Myr, 
with a peak of 0.015 M$_{\odot}$/yr.

\begin{figure*}
\begin{minipage}{\textwidth}
\centering
\includegraphics[width=0.49\textwidth,clip=]{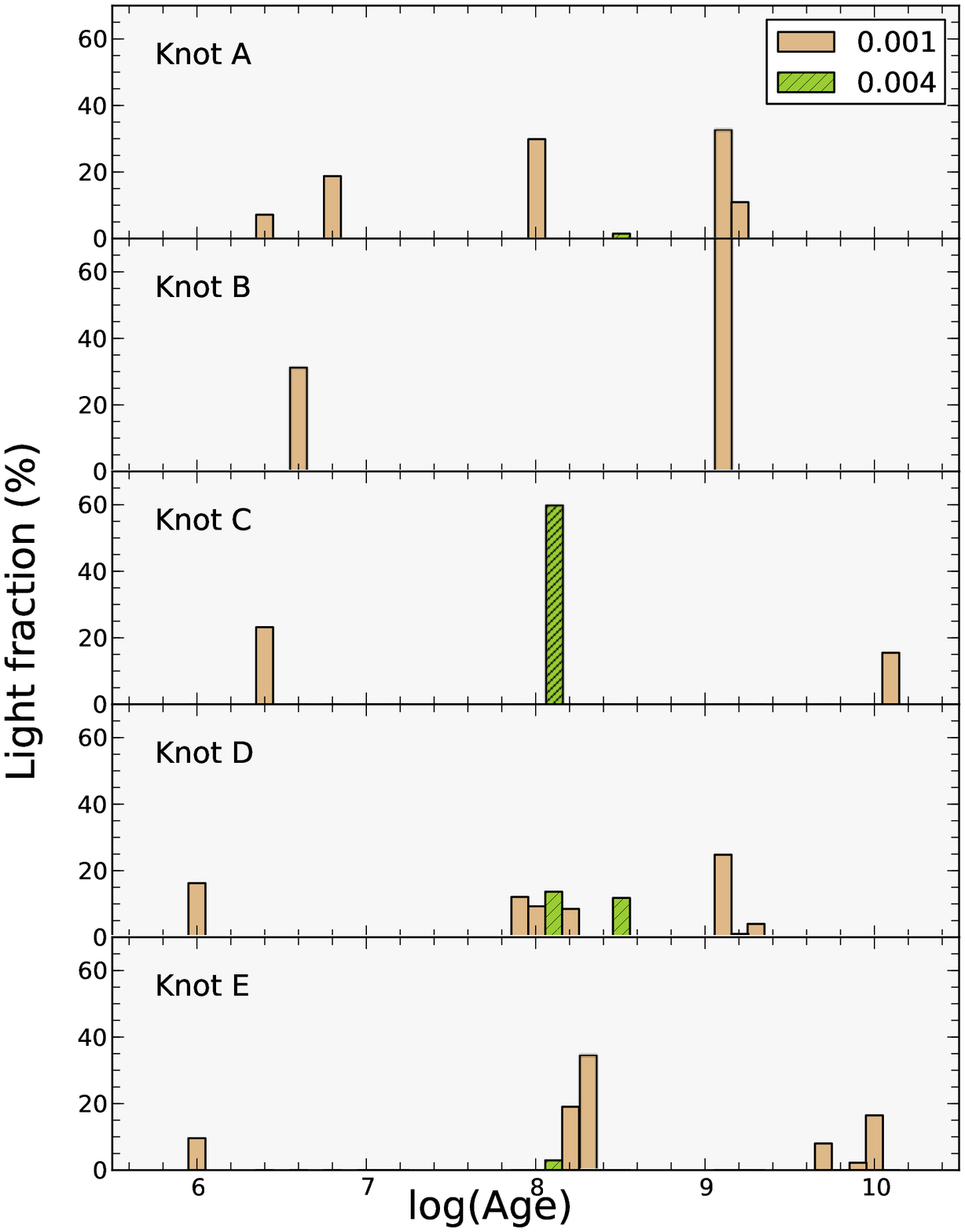}
\includegraphics[width=0.49\textwidth,clip=]{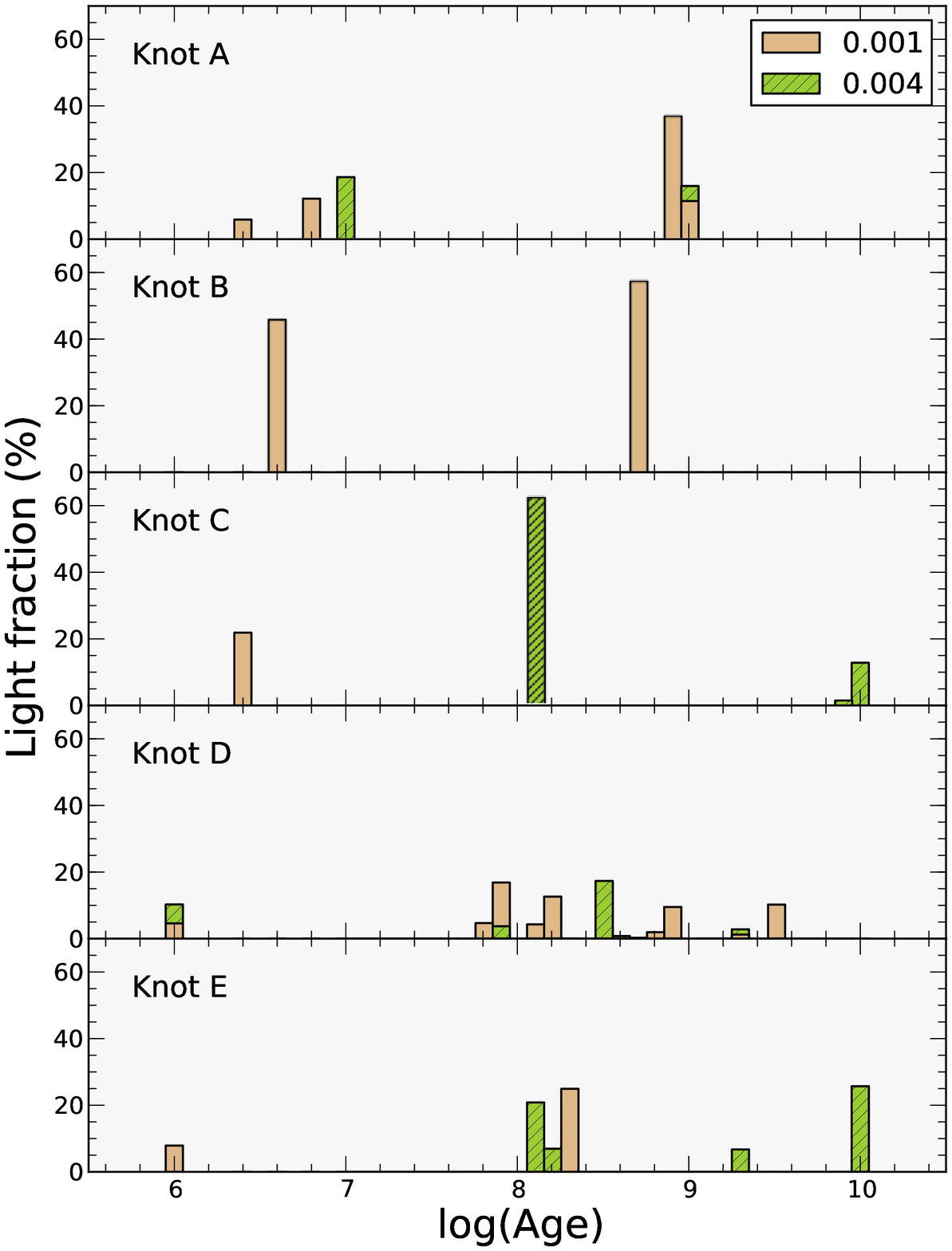}
\end{minipage}
\caption{Histograms of the distribution (light fraction) of the best fit derived by STARLIGHT 
for the spectra of the five knots of NGC 6789. The left panel corresponds to the fits 
using the constrains obtained by the CMD analysis, while the right panel 
shows the fits using non-constrained age-metallicity libraries.}
\label{sfhsl}
\end{figure*}

In the light of these results, we used the age-metallicity solution derived from the CMD 
as a prior to constrain the libraries of the STARLIGHT code, used to fit the spectra of 
the five knots (see Section \ref{line} for more details). In left panel of Figure 
\ref{sfhsl} we show the age distribution of the light fraction obtained from STARLIGHT for 
each knot under the constrains mentioned above. Another fitting was carried out without any 
specific age-metallicity constrain in order to compare it with the results of the first fitting. 
These are shown in the right panel of Figure \ref{sfhsl}. As can be seen, all knots present 
a very young stellar population, with ages younger than 10 Myr, responsible for the ionisation 
of the gas, combined with other older populations. The presence of these 
old populations is confirmed with the detection of absorption metal lines of 
Ca {\sc ii} K $\lambda$ 3933 \AA\ and Ca {\sc ii} H $\lambda$ 3968 \AA\, characteristic of 
old stellar populations, except for knot E, in which these features are marginally 
detected. The combination of these two different stellar population is also obtained by 
\cite{perez-montero10} in a sample of 10 BCDs using the same technique. In terms of mass, 
more than the 80\% comes from the older population.  The estimated total stellar mass and 
the fraction of the mass of the stellar population with an age younger than 10 Myr with 
respect to the total mass are given in Table \ref{slprop}. The stellar masses were corrected 
for aperture effects, with factors calculated using the ratio between the extinction-corrected 
H$\alpha$ fluxes measured in the elliptical regions of the photometry and those measured in 
the extracted long-slit observations for each knot. In the same table we show the EW(H$\beta$) 
values calculated once the contribution of the older stellar populations was removed, in order 
to use EW(H$\beta$) to derive the properties of the ionising stellar population (see 
Section \ref{stion}). The inner extinctions calculated by STARLIGHT for each fit are also 
listed in the same table. Finally, the results for the non-constrained case are given in the 
second row for each knot. The values from Table \ref{slprop} show that those fits obtained 
from non-constrained libraries do not significantly alter some of the properties of 
the stellar populations, such as the total stellar mass and the correction
factor for EW(H$\beta$), even that some relevant differences appear in the
age and metallicity distribution of the resulting SFHs. Therefore, we relied on the general 
chemical evolution solution derived by StarFISH to constrain the age-metallicity libraries of 
STARLIGHT. At any rate, the general distribution of the light is similar in both cases. We 
recall that the CMD SFH (STARFISH) is derived for the whole galaxy, while the spectra SFH 
(STARLIGHT) is derived for each individual knot. 

Less significant is the difference of the fitted spectra between both cases. For these two 
particular metallicities, the variation of the intensity of the first four Balmer lines on the 
residual spectra for each knot is well within the observational errors when using one or the 
other case. The change of age-metallicity distribution in the libraries only affects significantly 
in these two particular cases to the SFH in the assignments of the metallicity of 
the stellar populations, but the overall derived age distribution is very similar.

Regarding the extinction, we find some discrepancies between the values found by 
STARLIGHT and those derived using Balmer decrements. However, the average extinction value 
for all knots (0.52 $\pm$ 0.16) is very close to the CMD best fitting value given by StarFish 
(0.64 $\pm$ 0.08). It should be noted, though, that in principle it is not expected to find 
the same values in the gas and the stellar population. 

\begin{table}
\caption{Properties of the stellar populations as obtained with STARLIGHT for 
each knot of NGC 6789, including the inner extinction, the aperture corrected 
total stellar mass, the mass of stars younger than 10 Myr (M$_{ion *}$), and the 
EW(H$\beta$) corrected for the contribution of the older population. The first row 
for each knot corresponds to the constrained age-metallicity libraries, while the 
second row is for the unconstrained case.}
\label{slprop}
\begin{tabular}{l c c c c}
\hline
 ID &   A(V) & log M$_*$ &  log M$_{ion *}$ & EW(H$\beta$)$_c$ \\
    &  (mag) &                      &           &     (\AA)          \\
\hline
\hline
Knot A & 0.4 &  5.34 & 3.69 & 23 \\
       & 0.4 &  5.35 & 3.81 & 19 \\
Knot B & 0.3 &  5.26 & 3.16 & 30 \\
       & 0.7 &  5.22 & 3.68 & 30 \\
Knot C & 0.7 & 6.00 & 4.03 & 49 \\
       & 0.6 & 5.97 & 3.95 & 51 \\
Knot D & 0.5 &  5.42 & 3.79 & 24 \\
       & 0.5 & 5.51 & 3.66 & 24 \\
Knot E & 0.7 & 5.16 & 2.71 & 84 \\
       & 0.5 &  5.21 & 2.47 & 95 \\
\hline
\end{tabular}
\end{table}

\begin{table*}
\caption{Properties of the individual knots derived from tailored photoionisation models compared
with the observed values.}
\label{model}
\begin{tabular}{l  | c  c  | c  c  | c  c  | c  c  |  c  c }
\hline
                                &   \multicolumn{2}{c|}{NGC 6789-A}    &  \multicolumn{2}{c|}{NGC 6789-B}  & \multicolumn{2}{c|}{NGC 6789-C}         &         \multicolumn{2}{c|}{NGC 6789-D}  &  \multicolumn{2}{c}{NGC 6789-E}       \\
  & Mod. & Obs.  & Mod. & Obs. & Mod. & Obs. & Mod. & Obs. & Mod. & Obs. \\
\hline
\hline
log L(H$\alpha$) (erg/s)   &  37.32  &  37.23  $\pm$ 0.02 &  37.34 &  37.34 $\pm$ 0.02 & 37.51 &  37.50 $\pm$ 0.03 & 37.18 & 37.17 $\pm$ 0.02 & 37.03  &  37.02 $\pm$      0.01\\
Radius (pc)         &   7   &    30     &  33  &    34   &   22  &    37 &  10  &    36    &  5  &    21 \\
FUV-NUV (mag)         &  0.17 &  0.14 $\pm$ 0.16  &  0.28 & 0.28 $\pm$ 0.19  & 0.30 &    0.31 $\pm$  0.22 & 0.22 & 0.03 $\pm$ 0.18  &  0.10 & 0.06 $\pm$ 0.04 \\
I([O{\sc ii}]/I(H$\beta$)  &  2.30 & 2.34 $\pm$ 0.15 & 3.33 & 3.16 $\pm$ 0.20 & 2.35 & 2.51 $\pm$ 0.16 &
3.05 & 3.14 $\pm$ 0.20 & 1.09 & 1.06 $\pm$ 0.04 \\
I([O{\sc iii}]/I(H$\beta$)  &  3.82 & 3.76 $\pm$ 0.17 & 1.47 & 1.48 $\pm$ 0.07 & 3.45 & 3.44 $\pm$ 0.15 &
1.72 & 1.73 $\pm$ 0.08 & 2.30 & 2.26 $\pm$ 0.20 \\
12+log(O/H)   &   7.93  &   7.84 $\pm$ 0.04 &  7.82  & 7.93 $\pm$  0.10 & 7.79 & 7.81 $\pm$ 0.05 & 7.77 & 7.90 $\pm$ 0.12 &
7.64 & 7.80 $\pm$ 0.18 \\
-EW(H$\beta$ (\AA) &  21  & 23  & 30   &   30   & 48   & 49  &  29   &   24   &  80  &   84  \\
Age (Myr)   &   5.4   &   ...   &   4.2   &   ...   &   3.5   &    ...   &   5.2   &   ...   &   3.9 & ...  \\
Dust-to-gas ratio   &   0.026   &  ...     &    0.34   &  ...    &     0.12   &  ...  &    0.15   &  ... &  0.015  & ...  \\ 
Abs. factor ($f_d$)     &    5.34   &   ....   &  7.28  & ...  &   8.21   &    ...   &   4.37   &   ...  &    2.23 & ... \\
\hline
\end{tabular} 
\end{table*}

\subsection{Ionising stellar populations and photoionisation models}
\label{stion}

Among the properties of the young ionising populations that can be derived from 
optical spectroscopy are their stellar masses and ages. The former can be derived from
the H$\alpha$ luminosity and EW(H$\beta$) ({\em e.g.} \citealt{diaz98}).
H$\alpha$ luminosity can also be used to estimate the SFR and the mass of ionised hydrogen.  
The age can be estimated by using the EW(H$\beta$) as compared to sequences of evolutionary 
synthesis models. In our case, the contamination of the older underlying population to 
the H$\beta$ continuum was removed by using the analysis made with STARLIGHT. As can be seen 
in Table \ref{slprop}, these corrections are quite different between the knots, going from a 
3\% in the case of knot B to a 42\% in the case of knot E.

Nevertheless, as already shown by \cite{pmd07} and later in \cite{perez-montero10}
for other samples of BCDs, the corrected EW(H$\beta$) are still much lower than the
expected values for the ionisation of stellar clusters younger than 10 Myr. The 
H{\sc i} mass can be estimated using the CO observations of this object reported 
by \cite{leroy05} (marginally detected and thus, an upper limit) 
and using the $X_{CO}$ factor at the appropriate metallicity proposed by \cite{magrini11} to 
derive the H{\sc i} column density. For the adopted distance of NGC~6789 this gives a total
mass of log(M$_{HI}$/M$_{\odot}$) $\sim$ 6. Thus, this implies that dust absorption 
could be the main responsible of the disagreement between expected and observed EW(H$\beta$).
Thus, both L(H$\alpha$) and EW(H$\beta$) must be corrected using the dust absorption factor ($f_d$) 
before the derivation of the properties of the ionising cluster. This correction must be done even 
after the extinction correction, because it corresponds to the fraction of ionising photons 
absorbed by the dust and, therefore, not absorbed and re-emitted by the gas:

\begin{equation}
Q(H) = f_d \times Q_{obs}(H)
\end{equation}

To estimate both $f_d$ and the ages of the ionising cluster for each knot, we made tailored 
photoionisation models using the code CLOUDY v.08.00 \citep{ferland98}. Our models assume a 
one-dimensional structure, with the central ionising cluster and a gas and dust geometry, and 
compute the emergent spectrum. For each knot we took as input the Starburst99 libraries 
\citep{leitherer99,vazquez05}, based on stellar model atmospheres from \cite{smith02}, 
Geneva evolutionary tracks with high stellar mass loss \citep{meynet94}, a Kroupa IMF 
\citep{kroupa02} in two intervals (0.1-0.5 and 0.5-100 M$_\odot$) with different exponents 
(1.3 and 2.3 respectively), the theoretical wind model \citep{leitherer92}  
and a supernova cut-off of 8 M$_\odot$. We fixed the metallicity of the stellar 
populations to Z = 0.001 ($\approx$ 1/12 Z$_\odot$), which corresponds to the closest total 
oxygen abundance measured using the direct method in all the knots. We also fixed the density 
of particles to match the value derived in each knot from the [S{\sc ii}] emission-line ratio 
and the chemical abundances in the gas (except oxygen) to match those derived using the direct 
method. Those species that were not measured were put to scale with the solar proportions 
measured by \cite{asplund09}. 
Although not all species are expected to follow solar proportions, the deviation of this 
assumption does not affect the results of our models of all the other studied species.
In each knot, we used an iterative method to find the values 
of the free parameters (oxygen abundance, dust-to-gas ratio, number of
ionising photons, internal radius, filling factor, and age of the ionising cluster)
that better fit the observed properties (3727 {\AA} [O{\sc ii}]/H$\beta$,
5007 {\AA} [O{\sc iii}]/H$\beta$, 12+log(O/H), L(H$\alpha$), and FUV-NUV colour
index). The fitted properties along with the derived cluster age and $f_d$ in the 
corresponding model for each knot are shown in Table \ref{model}. 
The FUV-NUV colours were obtained by convolving the model output emergent spectrum for
each knot with the shape of the corresponding GALEX filter, performing
a direct measure of the resulting total flux, and finally calculating the
corresponding colour in AB magnitudes.

The assumptions taken for the models described above may be an
oversimplification as they do not take into account the three-dimensional structure
of the ionized gas and the dust. Besides, we cannot take as input parameter the 
dust-to-gas ratio of this galaxy and possible uncertainties related with the adopted
synthesis stellar atmospheres can affect the final result.

As can be seen, the observed properties are generally well fitted by the models, 
by assuming different geometries and dust-to-gas ratios without noticeably changing the 
derived metallicity of the gas.  
According to our results, $f_d$ correlates better with the FUV-NUV index 
that with the reddening constant derived from the Balmer decrement in the optical spectra. 
For instance, knot C has the highest $f_d$ with the lowest c(H$\beta$) and, in contrast, knots 
A and E have the lowest $f_d$ and the highest c(H$\beta$). This could be indicative of a 
complex inner dust structure which is not well traced by the reddening constant derived from 
the Balmer decrement. Regarding the cluster ages, our results indicate that all knots have
ionising cluster in a range of ages between 3.5 Myr for knot C and 5.4 Myr for knot A. The 
youngest cluster appears in knot C, although knot E, which has a very low $f_d$ correction,
has the lowest EW(H$\beta$). The age sequence in the studied knots does not correspond to any 
specific spatial order. 

\begin{table*}
\caption{Properties of ionising stellar populations in the knots as derived using the 
Q(H$^{0}$) value obtained from the photoionisation models.} 
\label{haprops}
\begin{tabular}{l c c c c c c c}
\hline
 ID   & log Q(H$^0$)  & log M$_{burst}$ & log M(H{\sc ii}) &  log SFS      & SFR \\
      &   (erg/s)     &   (M$_{\odot}$)  & (M$_{\odot}$) & (M$_{\odot}$) &  (10$^{-3}$ M$_{\odot}$/yr) \\    
\hline
\hline
Knot A          & 49.94 & 3.82  & 3.45     & 4.34 &  1.04  \\
Knot B          & 50.09 & 3.65  & 3.60     & 4.32 &  1.47  \\
Knot C          & 50.31 & 3.69  & 3.82     & 4.54 &  2.44  \\
Knot D          & 49.69 & 3.52  & 3.20     & 4.09 &   0.59  \\
Knot E          & 49.25 & 2.73  & 2.76     & 3.48 &   0.21  \\
NGC 6789$^{a}$  & 51.49 & $\cdots$ & 5.00 & 5.89 & 36.97  \\
\hline
\end{tabular}
\begin{center}
$^{a}$ Values derived from the H$\alpha$ image, not corrected for extinction. 
\end{center}
\end{table*}

The total masses of the ionising clusters were also derived by using the number of ionising 
photons and the ages obtained from the respective CLOUDY models and the Starburst99 
libraries of the corresponding synthesis cluster atmospheres. These are listed in 
Table \ref{haprops}. As can be seen, the stellar masses of the bursts in the four first 
knots are quite similar, while knot E has the lowest value. These masses are in agreement 
within 0.3 dex with the masses derived for the younger stellar population using the fitting 
to the optical spectrum made with STARLIGHT (Table \ref{slprop}). 

Once the $f_d$ were estimated for the five knots, it was possible to derive the 
properties that depend on the corrected total L(H$\alpha$), including
the mass of ionised hydrogen, the Star Formation Strengths (SFS, defined as the total 
mass of gas transformed into stars during the burst) and the SFR. These are listed in 
Table \ref{haprops}, along with the values for the whole galaxy.
The total ionised masses of the gas clouds in each knot were calculated 
using \cite{oster89}:

\begin{equation}
M_{H^+} = Q(H^{0}) \frac{m_{p}}{n_{e}\alpha_{B}}
\end{equation}

\noindent assuming case B, n$_{e}$ = 100 cm$^{-3}$, and 
$\alpha_{B} \sim 2.59 \times 10^{-13}$ cm$^{-3}$ s$^{-1}$.
The individual knots show masses of ionised hydrogen of the order of hundreds of solar masses 
for knot E and up to $\sim$ 6600 M$_{\odot}$ in knot C.
In the case of the whole galaxy, the total flux described in Section \ref{Halfa} can be 
converted into (1.0 $\pm$ 0.1) $\times$ 10$^{5}$ M$_{\odot}$ of ionised hydrogen, comparable to 
knot A of the Giant Extragalactic H{\sc ii} Region NGC 5471 in M101 \citep{garcia-benito11} or 
to the whole NGC 604 giant H{\sc ii} region in M33 \citep{relano09}. 

It is known that the SFR determined from H$\alpha$ is sensitive to several uncertainties, 
as those related with the extinction and the assumed IMF. Besides, not always  
all the ionising photons are absorbed, being the escape fraction of ionising radiation from 
individual H{\sc ii} regions (in nearby galaxies) between 15\% and 50\% 
\citep{kennicutt98}. 
We applied the calibration by 
\cite{oti-floranes10}\footnote{Their web tool can be found at\\ 
\texttt{http://www.laeff.cab.inta-csic.es/research/sfr/}}, which makes a distinction 
between instantaneous (IB) and extended burst (EB). We used our extinction and dust absorption 
corrected H$\alpha$ luminosity (except for the value of the whole galaxy) and a Salpeter IMF in 
the range 0.1-100 M$_{\odot}$ as input. For the IB case, the value given by the 
calibration is the SFS. The SFS values in Table \ref{haprops} were calculated using the age 
provided by the calibration, namely 4, 5, and 6 Myr, closest to the age estimated from the 
photoionisation models. According to this calibration, the mass of gas transformed in stars 
during the actual burst in each knot is of the order of tens of thousands of solar masses, 
except for knot E, which is one order of magnitude lower.

We provide as well the SFR, for the EB case (continuous star formation), as a reference value. 
Regarding the SFR of the whole galaxy, the value estimated by means of the H$\alpha$ flux is
slightly higher than the peak obtained from the CMD (see Figure \ref{sfh}). 
The SFR for BCDs spans from a few 10$^{-3}$ to several times 10$^{1}$ M$_{\odot}$/yr 
\citep{hopkins02,zhao11}, showing NGC 6789 not a particularly high value.

Although our models are able to reproduce much of the observational
information both in the optical and the UV, some of their limitations must be
taken into account. It is possible that the EW(H$\beta$) could be modified owing to great 
fractions of leaked photons. However, comparing the mass of ionised gas in this galaxy 
[log(M/M$_{\odot}$) $\sim$ 5, see Table \ref{haprops}] with the mass of neutral hydrogen 
does not support this scenario, but taking into account the oversimplified geometry of our 
models it cannot be ruled out completely. In this sense, as the long-slit 
does not cover entirely the gas-emission from the knots, the 
EW(H$\beta$) are lower limits. Nonetheless, the aperture factors 
derived from the direct comparison with the H$\alpha$ image do not 
lead to a satisfactory solution.

Larger dust-to-gas ratios and $f_d$ than typical values for H{\sc ii} regions 
with larger metallicity ({\em e.g.} \citealt{inoue01}) were derived by the models 
in all knots in order to fit the observed EW(H$\beta$) and FUV-NUV colours.  
However, these values cannot be taken as representative of the whole galaxy
but only for the analysed star-forming knots, considering that the fraction of 
ionised-to-neutral gas is quite small, as mentioned above, and that this
galaxy does not present a very high IR luminosity ({\em i.e.} it does not
appear in the IRAS catalogue). We have run other set of models with  
standard dust-to-gas ration values, but they do not fit the observables mentioned 
above. The dust-to-gas ratios could be considerably decreased by assuming 
different geometries of the gas and dust inside the H{\sc ii} regions, but 
this approach is beyond the scope of this work.

Finally, our results about the masses and ages of the ionising clusters must be used 
only as a guiding signpost, as the involved masses range in a regime 
($\sim$ 10$^4$ M$_{\odot}$) where stochastic fluctuations in the IMF can be important 
\citep{cervino03,cervino06}. In this sense, IMF sampling effects 
\citep{pflamm-altenburg07,pflamm-altenburg09} can also alter our results, as there can be 
a lower number of massive stars than expected for the mass range of these knots.

\subsection{Densities, temperatures and chemical abundances}

The derived densities, both for [S{\sc ii}] and [O{\sc ii}], are very similar for the first 
four knots, being lower than 100 cm$^{-3}$. Interestingly, knot E, located in the outskirts
of the galaxy, shows a density twice higher than the knots located in the central part of 
the galaxy. At any rate, these densities are typical of low density environments found in BCDs.

Regarding electron temperatures, T([O{\sc iii}]) was measured in the five knots,  with errors 
which depend on the quality of the spectrum in each knot going from 3\% in knot A up to 13\% in 
knot E. The first four star-forming regions show very similar [O{\sc iii}] temperatures, all 
within a relatively narrow range, between 13000 and 14300 K, while knot E has a lower 
temperature by about 2000 K. To our knowledge, there is no previous report on T([O{\sc iii}]) 
for this galaxy in the literature.

T([O{\sc ii}]) was also measured in three of the knots (B, C, and E) and it was estimated in 
the other two knots by means of relations between these temperatures and the measured
value of T([O{\sc iii}]). The knot C, 
which is the brightest in H$\alpha$, has in average higher temperatures.

We derived oxygen, nitrogen, sulphur, neon and argon total abundances in all the 
studied knots by taking the ionic abundances calculated in Section 
\ref{abundances} and the ICFs estimated in our tailored photoionisation models. 
As in the case of temperatures, no other direct abundance determination exists in 
the literature for this galaxy. In average, NGC 6789 shows an oxygen abundance 
characteristic of the low values found in strong line BCDs, with values in all knots 
in the range of 12 + log(O/H) = 7.80-7.93, what is compatible with a similar abundance 
in all knots taking into account the errors. This behaviour, i.e. showing 
different star forming knots within a BCD close abundance values (i.e., $<$ 0.2 dex), 
is also found in other objects using both long-slit spectroscopy 
\citep{papaderos06,cairos09,perez-montero09a,hagele11} 
and integral field spectroscopy \citep{kehrig08,perez-montero11}, or 
in H{\sc ii} complexes in spiral galaxies with a similar spatial scale 
\citep{kennicutt03,garcia-benito10}.

Since nitrogen and oxygen have different nucleosynthetic origins, their ratio is related to the 
chemical history of galaxies. Primary nitrogen synthesis is thought to occur in intermediate-mass 
stars in the CNO process during hydrogen burning being hence independent of the initial 
heavy-element abundances, while secondary nitrogen production is expected to be produced in 
stars of all masses \citep{vilacostas93}. At low metallicity, most part of nitrogen has 
a primary origin and a constant log(N/O) ratio is observed. However, the N/O values found 
in the studied knots of NGC 6789 range in an interval (from -1.38 to -1.26) is sensibly 
higher than the observed values for other BCDs or low-metallicity dwarf irregular 
galaxies \citep{izotov99a,vilchez03,vanzee06,pmc09}, which is around log(N/O) 
$\approx$ -1.6. In the case of knot C, this ratio is even higher, with
a value of -1.07.  
The cause of these relatively high N/O can be partially found
in the corresponding ICFs derived by the models as
the total nitrogen abundances derived from the assumption N/O $\approx$ N$^+$/O$^+$
give lower N/O values in knots A [log(N/O) = -1.41] and C
[log(N/O) = -1.34], although still higher within the errors than the 
typical N/O value for BCDs.

On the other hand, several causes are cited in the literature to explain this
overabundance of nitrogen in low-metallicity environments ({\em e.g.} the pollution
of the ISM by Wolf-Rayet stellar winds), but it is especially indicative in this case that
we found this overabundance in the five knots, which present different evolutionary
and excitation properties. Therefore, the high N/O is apparently related with some other process
affecting all the ISM of the galaxy. The fall of pristine gas in the galaxy could explain
the N/O overabundance \citep{koppen05} and, at same time, the triggering of the star 
formation in different places of the galaxy with very short time intervals between them.

The ratio of the alpha elements, sulphur, neon, and argon, to oxygen should be constant and show 
no dependence on the oxygen abundance, since all are products of $\alpha$-processes in the 
same massive stars that make oxygen. The derived log(S/O) ratios range between 
-1.64 and -1.41 in the four first knots and it is sensibly higher in knot E, but
in this knot the derivation is much more uncertain as no [S{\sc iii}] was measured. 
Although the ratio seems to increase from knot A to D, the average error ($\sim$ 0.18) 
prevents us from drawing any conclusion about the homogeneity.

Regarding the logarithmic Ne/O ratio, knots A and C show the same value, -0.70, while for 
the rest of the knots the mean value is -0.94, with an average error of 0.07 dex for A and C 
and 0.20 dex for B, D, and E.

The Ar/O ratios found, in the range -2.28 to -2.06, show a very similar value for 
all knots, taking 0.17 dex as the average error (ranging from 0.09 up to 0.22 dex).

Finally, the derived helium abundances are the same for all five knots of NGC 6789 within 
observational errors, and similar to the values found for other BCDs \citep{izotov04c,hagele08}.

\section{Conclusions}
\label{conclusions}

In this paper, we present resolved stellar, H$\alpha$, and GALEX photometry, 
from different data archives, and WHT-ISIS optical spectroscopy of the five brightest 
star-forming knots of the nearest BCD galaxy NGC 6789 in order to study their SFH and 
metallicity.

The spectroscopic observations of NGC 6789 were taken using ISIS double-arm spectrograph 
attached to the 4.2m WHT, which allowed the simultaneous analysis of the spectra, covering 
from 3650 up to 7500 \AA. From the long-slit spectra, we extracted and performed a detailed 
analysis of the five main star forming knots of the galaxy. Thanks to the measure of the 
[O{\sc iii}] electron temperature in all knots and the measure of T([O{\sc ii}]) in three 
of them, we provide ``direct method'' abundances for oxygen, nitrogen, sulphur, neon,
and argon, not reported previously in the literature. Our analysis indicates that this 
galaxy is metal-poor (12+log(O/H) in the range 7.80 - 7.93) and chemically homogeneous, 
with quite similar values of the studied species in all knots. At same time, all the 
knots present values of N/O which are higher than expected for the metal regime of this 
galaxy.

We used optical data obtained from the HST archive to derive the SFH of NGC 6789 by 
means of the classical method of CMD reconstruction. 
We used the derived SFH in the program STARLIGHT to fit the optical spectrum
and we compared these results with another non-constrained case.
We corrected the emission-line measures from absorption and we derived the
total stellar masses for each knot.  Although the SFH obtained from
the CMD is not foreseen by STARLIGHT in the non-constrained case,
we checked that the total stellar mass and the ratio of older and younger
stellar populations are not noticeably affected by the assumed SFH.
Anyway, whenever possible, the extension of this type of
studies for objects for which simultaneous resolved stellar photometry
and optical spectroscopy are available will provide a much more accurate
constrain to the age-metallicity-extinction degeneracy of these galaxies, as
well as the comparison between nebular and stellar properties.

Finally, the ages and masses of the bursts of star formation in each knot were derived 
using CLOUDY tailored photoionisation models to fit the optical and UV photometric 
properties and the observed emission-line ratios and corrected H$\beta$ equivalent widths.
We found that dust absorption factors correlate much better with the GALEX FUV-NUV 
colour index than with the reddening constants derived from the Balmer decrement, 
indicating a very complex inner dust extinction structure. The models predict for all 
knots instantaneous bursts with ages in the range between 3 and 6 Myr. These ages do 
not follow any spatial trend in the galaxy image, so they are possibly related with 
the distribution of the galaxy in the line of sight. The dust absorption-corrected 
H$\alpha$ fluxes were used to derive accurate SFRs for the individual knots.

The combination of several observational and model techniques lead to a better and auto-consistent 
study of NGC 6789. The derivation of a non-typical metallicity evolution using a CMD allows the use 
of the SFH to the subtraction of the older stellar population using spectral fitting to the optical 
spectrum with STARLIGHT. This information, together with the derivation of accurate physical properties 
and ionic chemical abundances pointing to similar low O/H and high N/O ratios in all knots allow the 
application of photoionisation models that predict dust-absorption factors fitting the GALEX colour 
indices and the derivation of ages (IBs in the range 3-6 Myr). 

\section*{Acknowledgements}

R.G.B. acknowledges support from the China National Postdoc Fund Grant No. 20100480144 and 
MICINN AYA2010-15081. This work has been partially supported by DGICYT grant AYA2007-67965-C03, 
AYA2007-67965-C03-02, AYA2007-64712, and Junta de Andaluc\'ia TIC 114. 
The WHT is operated on the island of La Palma by the ING in the Spanish
Observatorio del Roque de los Muchachos of the Instituto de Astrof\'isica de
Canarias. We thank the Spanish allocation committee (CAT) for awarding
observing time. We thanks We would like to thank to Armando Gil de Paz, for allowing the 
study of B, R, and H$\alpha$ images of NGC~6789 in the Palomar/Las Campanas Atlas 
of Blue Compact Galaxies. We thank Roberto Cid Fernandes and the people of the 
STARLIGHT Project Team (UFSC, Brazil), for making the STARLIGHT code available. 
We acknowledge fruitful discussions with Enrique P\'erez and Ricardo Amor\'in. We 
thank an anonymous referee for very useful comments that improved 
the presentation of the paper.

\end{document}